\documentclass[10pt,journal,compsoc]{IEEEtran}
\ifCLASSOPTIONcompsoc
  \usepackage[nocompress]{cite}
\else
  \usepackage{cite}
\fi
\ifCLASSINFOpdf
  \usepackage[pdftex]{graphicx}
  \graphicspath{{./figs/}}
  \DeclareGraphicsExtensions{.pdf,.jpeg,.png}
\else
\fi
\usepackage{amsmath,amssymb,amsfonts}
\usepackage{algorithmic}
\usepackage{array}
\ifCLASSOPTIONcompsoc
  \usepackage[caption=false,font=footnotesize,labelfont=sf,textfont=sf]{subfig}
\else
  \usepackage[caption=false,font=footnotesize]{subfig}
\fi
\usepackage{url}
\newcommand\MYhyperrefoptions{bookmarks=true,bookmarksnumbered=true,
pdfpagemode={UseOutlines},plainpages=false,pdfpagelabels=true,
colorlinks=true,linkcolor={black},citecolor={black},urlcolor={black},
pdftitle={CCGL: Contrastive Cascade Graph Learning},%<!CHANGE!
pdfsubject={Computer Science},%<!CHANGE!
pdfauthor={Xovee Xu, Fan Zhou, Kunpeng Zhang, and Siyuan Liu},%<!CHANGE!
pdfkeywords={Information cascade, contrastive self-supervised learning, graph learning, information diffusion, data augmentation.}}%<^!CHANGE!
\ifCLASSINFOpdf
\usepackage[\MYhyperrefoptions,pdftex]{hyperref}
\else
\usepackage[\MYhyperrefoptions,breaklinks=true,dvips]{hyperref}
\usepackage{breakurl}
\fi
\newtheorem{definition}{\bf Definition}

\newcommand{\M}{CCGL}

\usepackage{booktabs}

\usepackage[table,xcdraw]{xcolor}
\usepackage{mdwmath}

\usepackage{algorithm}
\usepackage{algorithmic}
\usepackage{multirow}
\usepackage{dsfont}
\usepackage{pifont}

\usepackage{enumitem}

\newcommand{\vpara}[1]{\vspace{0.03in}\noindent\textbf{#1}}

\begin{document}

\title{CCGL: Contrastive Cascade Graph Learning}


% ~\IEEEmembership{Member,~IEEE,}
%         John~Doe,~\IEEEmembership{Fellow,~OSA,}

\author{Authors}
\author{
Xovee~Xu,~Fan~Zhou,~Kunpeng~Zhang,~and~Siyuan~Liu
\IEEEcompsocitemizethanks{\IEEEcompsocthanksitem Xovee~Xu and Fan~Zhou are with University of Electronic Science and Technology of China, Chengdu, Sichuan 610054 China. \protect\\E-mails: xovee@ieee.org, fan.zhou@uestc.edu.cn.
\IEEEcompsocthanksitem Kunpeng~Zhang is with the University of Maryland, College park, MD 20742 USA. E-mail: kpzhang@umd.edu.
\IEEEcompsocthanksitem Siyuan~Liu is with the Pennsylvania State University, PA 16802 USA. E-mail: siyuan@psu.edu.
}
\thanks{Manuscript received 9 Dec. 2020; revised 20 Dec. 2021; accepted 7 Feb. 2022. Date of publication 0 .0000; date of current version 0 0.0000. \protect\\ This work was supported in part by the National Natural Science Foundation of China under Grants 62072077 and 62176043. 
\protect\\ (Corresponding author: Fan Zhou.)\protect\\
Recommended for acceptance by X. Xiao\protect\\
Digital Object Identifier no. 10.1109/TKDE.2022.3151829}
}

\markboth{IEEE Transactions on Knowledge and Data Engineering}%
{Xu \MakeLowercase{\textit{et al.}}: CCGL}
\IEEEtitleabstractindextext{%
\begin{abstract}
Supervised learning, while prevalent for information cascade modeling, often requires abundant labeled data in training, and the trained model is not easy to generalize across tasks and datasets. It often learns task-specific representations, which can easily result in overfitting for downstream tasks. Recently, self-supervised learning is designed to alleviate these two fundamental issues in linguistic and visual tasks. However, its direct applicability for information cascade modeling, especially graph cascade related tasks, remains underexplored. In this work, we present Contrastive Cascade Graph Learning (\textbf{CCGL}), a novel framework for information cascade graph learning in a \textit{contrastive}, \textit{self-supervised}, and \textit{task-agnostic} way. In particular, CCGL first designs an effective data augmentation strategy to capture variation and uncertainty by simulating the information diffusion in graphs. Second, it learns a generic model for graph cascade tasks via self-supervised contrastive pre-training using both unlabeled and labeled data. Third, CCGL learns a task-specific cascade model via \textit{fine-tuning} using labeled data. Finally, to make the model transferable across datasets and cascade applications, CCGL further enhances the model via \textit{distillation} using a teacher-student architecture. We demonstrate that CCGL significantly outperforms its supervised and semi-supervised counterparts for several downstream tasks. 
\end{abstract}

}

\maketitle

% is not selected <OR> if conference mode is selected - because compsoc
% conference papers position the abstract like regular (non-compsoc)
\IEEEdisplaynontitleabstractindextext

% \fi
\IEEEpeerreviewmaketitle

\ifCLASSOPTIONcompsoc
\IEEEraisesectionheading{\section{Introduction}\label{sec:introduction}}\IEEEPARstart{I}{n} recent years, information cascades have received considerable attention in various research areas, such as decision support systems, social network analysis, and graph learning~\cite{zhou2021survey}. Understanding cascades becomes important and can lead to significant economical and societal impacts. For example, predicting the number of affected cases and deaths as well as the ``superspreaders'' in a region during the COVID-19 pandemic is critical for policy makers to plan subsequent actions \cite{chang2021mobility}. Such a predictive task typically involves several key modeling components. Specifically, the number of affected cases in one region (e.g., county) might be very related to its neighbors in a geo-network at a region-level, since the mobility is an important factor of COVID-19 transmission. This in turn indicates that a good network modeling and representation might affect the prediction performance. This forecasting task also often involves modeling dynamics, e.g., the number of affecting cases evolves and depends on the past. In addition, the size of (labeled) data could be limited, which requires extra effort to model variation and uncertainty to learn generic knowledge in data and thus alleviate overfitting and enhance knowledge transfer. 
With the advances of deep neural networks, many supervised learning approaches for modeling information cascades have been proposed. For example, recurrent neural networks (RNNs) were applied to model users in cascade and their temporal dependencies \cite{yang2021full}; deep language models and convolutional neural networks are implemented to learn words and visual representations for the content in cascades; and graph embedding techniques and graph neural networks (GNNs) are usually used to learn diffusion structures \cite{li2017deepcas,tang2021fully}.

However, being supervised, the model training of these approaches requires abundant labeled data which is expensive to obtain~\cite{cao2017deephawkes}. And these models are not easily generalized across datasets and different cascade-related applications. Furthermore, a large amount of unlabeled data exist and are often not utilized in these models. To leverage unlabeled data for learning more generic representations, researchers have developed various unsupervised or semi-supervised methods, primarily focusing on the generative models, such as unsupervised domain adaptive graph convolutional network (GCN)~\cite{wu2020unsupervised}, auto-encoding variational Bayes~\cite{kingma2013auto}, and strategies for pre-training GNN~\cite{hu2020strategies}. These methods intend to learn how to reconstruct instances via embedding low-level features in a fine-grain detailed manner, which can easily lead to overfitting \cite{cao2017deephawkes}. 
In many situations (e.g., information cascade prediction), we need to learn general knowledge from both labeled and unlabeled data without relying solely on specific downstream task and/or label supervision.
That being said, learning such representations in an abstractive way to capture high-level semantics might be more useful, e.g.: (i) a more generalized model to different tasks and data; (ii) the characteristics of unlabeled data can be utilized; (iii) the model's performance can be improved via fine-tuning and distillation; and (iv) better transfer the learned knowledge to other prediction tasks and datasets. 
These benefits, in turn, motivate us to consider contrastive self-supervised learning. 
Such learning paradigms have made significant advances in both natural language processing (NLP) and computer vision (CV), especially shedding more light on improving learning ability of models without human supervision and model generalization via knowledge transfer. However, its applicability of understanding information cascade on graphs still remains underexplored in the community. 

Directly applying contrastive self-supervised learning to graph-based information cascade tasks is desired to study but faces several big challenges, such as: 
(i) how to learn generic knowledge of graph cascades in a contrastive, self-supervised, and task-agnostic way, in particular how to leverage a large amount of unlabeled cascade data; 
(ii) how to construct positive and negative sample pairs in a contrastive learning framework while capturing variations of data and the dynamic diffusion characteristics of cascades;  
and (iii) how to fine-tune the pre-trained model for downstream cascade prediction tasks in a semi-supervised and task-specific way. In addition, how to distill the model for knowledge transfer across applications and datasets -- while mitigating the effect of ``negative transfer'' -- is another hurdle to overcome. 

To tackle the above obstacles, we introduce a general semi-supervised learning framework, \textbf{\M}~(\underline{C}ontrastive \underline{C}ascade \underline{G}raph \underline{L}earning), in which information cascade graphs are augmented by simulating the information diffusion in graphs: manually perform perturbations (\textit{adding} and \textit{removing} actions on both nodes and edges), and manipulate node features. \M~does not require label information for pre-training, and the trained model is further fine-tuned and distilled for downstream cascade prediction tasks and different datasets. 
The pre-trained model focuses on learning the inner differences and characteristics between cascade graphs rather than solely for accurate prediction, providing a better starting point for supervised training.
A comparison between traditional supervised models and \M~is shown in \figurename~\ref{fig:teaser}. To summarize, the main contributions of our work are as follows:

\begin{itemize}[leftmargin=*]
    \item To the best of our knowledge, this is the first work to utilize unlabeled cascades, devise an effective data augmentation strategy, and design contrastive self-supervised learning for general cascade modeling and prediction. We further employ the semi-supervised fine-tuning and model distillation to improve the cascade prediction. 
    \item We propose a novel framework \M\footnote{Five datasets, pre-trained and fine-tuned models, as well as source codes are publicly available at \url{https://github.com/Xovee/ccgl}.} which learns cascade graph representations in two steps: (i) with labeled/unlabeled cascades and graph data augmentation, we pre-train the model and learn cascade representations in a \textit{self-supervised} and \textit{task-agnostic} way. In particular, we create different cascade graph views and use a contrastive loss to discriminate between similar and dissimilar cascade graphs; and (ii) we fine-tune and distill the pre-trained model in a \textit{semi-supervised} and \textit{task-specific} manner.
    \item Extensive experiments on five real-world datasets are conducted to show \M's effectiveness, robustness, and generalizability compared to supervised counterparts.
\end{itemize}

With all the new designs together, \M~achieves the state-of-the-art performance in information cascade graph prediction, and we have several interesting findings: 
(i) instead of potential overfitting, \M~uniformly improves prediction performance on all five cascade datasets with fewer labeled data, e.g., with only 1\% of labeled cascades, fine-tuned and distilled \M~decreases the cascade popularity prediction errors up to 9.2\%, and improves the cascade outbreak prediction accuracy up to 19.9\%; 
(ii) larger models and deeper MLP-based projection heads are essential for cascade self-supervised learning, while the larger batch sizes and longer pre-training epochs do not bring additional improvements; 
(iii) a teacher-student distilling framework is essential to the usage of unlabeled data and mitigates the negative effect of knowledge transferring; and (iv) for the knowledge transferring across datasets and two cascade prediction tasks, \M~outperforms supervised counterparts by non-trivial margins. 

The remainder of this paper is organized as follows. In Section~\ref{sec:related-work} and \ref{sec:preliminaries}, we give a detailed literature review of related work and necessary preliminaries.  Section~\ref{sec:methodology} presents the fundamentals of our proposed framework \M. In Section~\ref{sec:experiments}, we evaluate \M~on five large-scale information cascade datasets (Weibo, Twitter, ACM, APS and DBLP) and two downstream tasks (popularity prediction and outbreak prediction). Furthermore, we conduct extensive ablation studies and sensitive analyses. Section~\ref{sec:conclusion} concludes this work and points out potential future directions.

\begin{figure}[t]
    \centering
    \includegraphics[width=\linewidth]{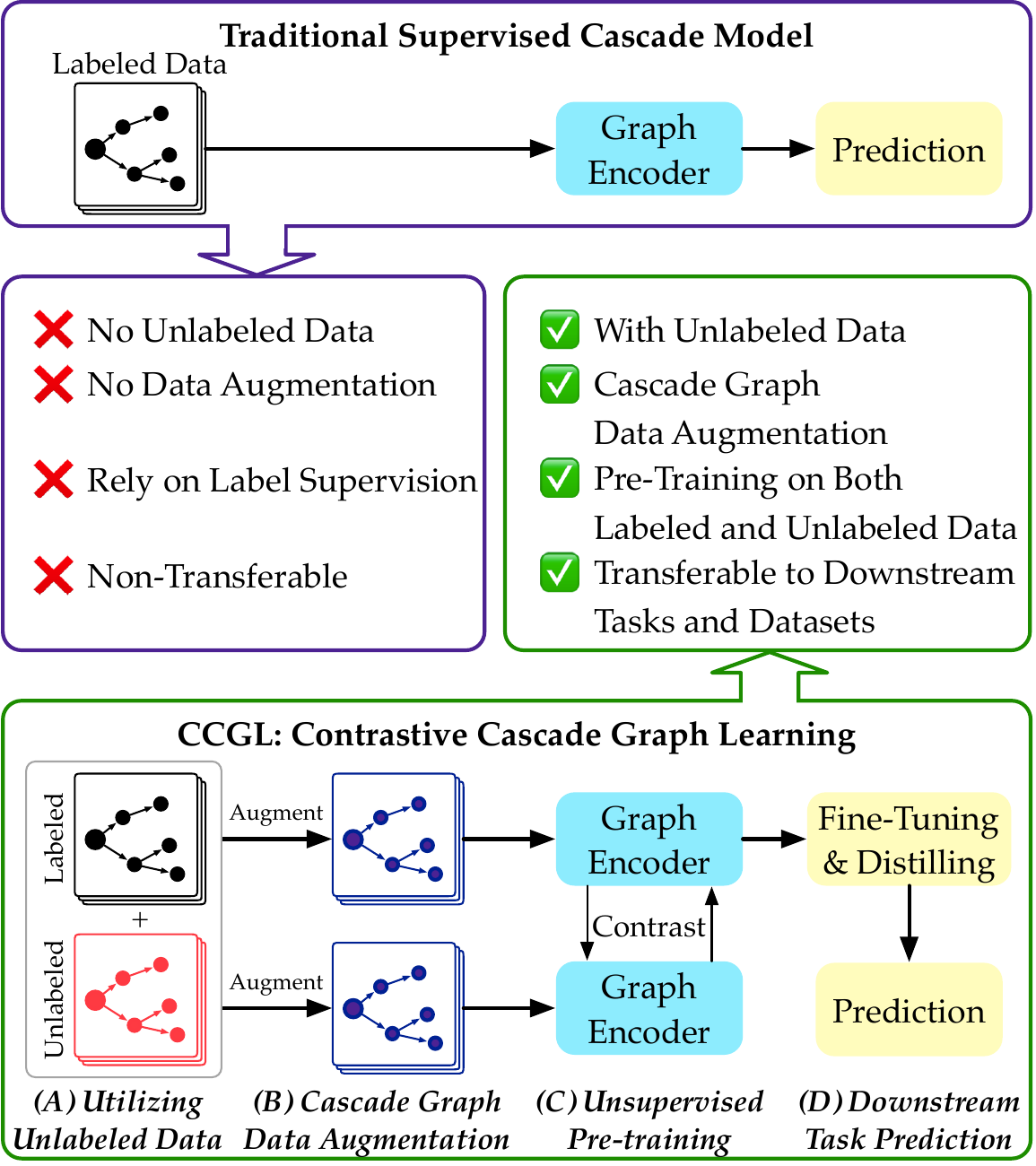}
    
    \caption{
    An illustration example of \M~compared to traditional supervised cascade learning models.
    }
    \label{fig:teaser}
\end{figure}

\else
\section{Introduction}
\fi

\section{Related Work}
\label{sec:related-work}
We overview four main streams of relevant literature and position our contribution in that context. 

\vpara{Information cascade modeling.}
This has been a long-standing and critical problem in the field of information diffusion and social network analysis \cite{zhou2021survey}. Most of existing works on information cascades can be categorized into three distinct groups:

\begin{itemize}[leftmargin=*]
    \item \textit{Feature-based models} characterize information cascades in different aspects by hand-crafted features, e.g., temporal and structural characteristics, textual and visual content, metadata and historical behaviors \cite{cheng2014can,zhao2018comparative}.

    \item \textit{Temporal models} mainly study the time-series data, incorporated by additional social information. They adopt various stochastic processes to model and simulate the information diffusion in networks, in a statistical and generative manner \cite{zhao2015seismic,gao2020using}; 

    \item \textit{Deep learning-based models} are utilizing state-of-the-art techniques from neural networks to learn expressive representations of cascades \cite{zhou2020variational,cao2020popularity}.
    Among them many adopted techniques from recurrent and graph neural networks. 
    Although they have achieved promising results in comparison to more traditional approaches, relying on supervised models still need extensive annotated labels and lack of generalization capability. 
\end{itemize}

\vpara{Self-supervised learning (SSL)} 
leverages data itself as supervision and therefore exploits the massive unlabeled data for improving representation learning and downstream tasks. Existing SSL methods usually follow a paradigm called contrastive learning, which learns representations by contrasting positive and negative instances. 
For example, discriminative approaches have been proposed, mainly studying the design of pretext tasks and contrastive losses:

\begin{itemize}[leftmargin=*]
    \item \textit{Pretext tasks} are generally for recovering noisy data, predicting neighboring words, or transforming the original data. Examples include instance discrimination between positive and negative samples \cite{he2020momentum,wu2018unsupervised}, global-local contrast for relative position prediction \cite{misra2020self} or mutual information maximization \cite{bachman2019learning,velickovic2019deep,oord2018representation}. 
    These pretext tasks are used to learn and extract useful data representations, but not for actual demands.
    \item \textit{Contrastive losses} focus on the similarity between positive and negative samples and are critical for learning good representations, along with negative sampling strategies. 
Previous work have come up with many learning mechanisms, e.g., end-to-end \cite{bachman2019learning}, memory mechanism \cite{he2020momentum}, projection head, augmentation and fine-tuning \cite{chen2020big,chen2020simple}.
\end{itemize}

\vpara{Data augmentation on graphs.} 
One of the critical components in SSL is data augmentation, used to improve the capability of model generalization and performance. Augmentation procedures are used in NLP and CV domains~\cite{cubuk2019autoaugment}
and previous works have investigated various strategies for data augmentations~\cite{chen2020simple,misra2020self}.
Unlike texts and images, graph data is usually non-Euclidean, sparse, and complex. Graphs can be directed, dynamic, attributed, and heterogeneous -- which further increases the complexity of their modeling. 
Since graph data have no analogous augmentations within languages and images (due to their irregularities), there are very few works that tackle the graph data augmentation problem. How to define an effective graph augmentation strategy is a challenging and non-trivial task to retrieve general knowledge from graph structures under an unsupervised learning paradigm.  Existing augmentation procedures for images, such as rotation, random crop and resize, cutout, color distortion, Gaussian blur, etc.~\cite{chen2020big}, cannot be directly applied to graph data for pre-training and fine-tuning~\cite{hu2020strategies}.

The most straightforward way to augment graph data is adding or removing nodes and/or edges. 
However, such operations may confront multiple obstacles~\cite{zhao2020data}, including: how to choose the target nodes/edges to add or remove, how to label the newly added nodes/edges, how to process the features associated with these nodes/edges, etc. 
There are few attempts in the existing literature, such as removing edges at random and covering node features~\cite{rong2019dropedge,hu2020strategies} to prevent over-fitting and over-smoothing; relieving the over-smoothing for GNNs by adding or removing edges between nodes based on model predictions~\cite{chen2020measuring}; utilizing graph auto-encoder as an edge predictor module to add ``missing'' edges and remove ``noisy'' edges \cite{zhao2020data}.
GCC designs a pre-training task as a sub-graph instance discrimination. An $r$-ego network was augmented by random walk with restart, sub-graph induction, and anonymization, to capture general and transferable patterns in the graph structure \cite{qiu2020gcc}. 
Augmenting graphs by manipulate (i) node features, e.g., masking or adding Gaussian noise; and (ii) graph structures, e.g., adding/removing connectivities or sub-sampling \cite{hassani2020contrastive}.

In \M, we design a new data augmentation strategy specifically for information cascade graph. 
Different from existing graph data augmentations which cannot be directly used for cascade graphs (as we will explain in Section \ref{subsec:4.2}), we simulate the information diffusion process to create new graph views as re-diffusion, and then contrast them for cascade representation learning.
The unique aspects of our augmentation strategy are: 
(i) they are suitable for cascade graphs, while most of the previous models are not; 
(ii) we add and remove both edges and nodes, as well as features in the graph; and (iii) we do not study node/graph classification but information cascade prediction.

\vpara{Pre-training and transfer learning on graphs.} 
One important application of unsupervised graph pre-training is to learn a transferrable graph encoder for downstream graph-based tasks. While commonly seen in CV and NLP domains, there are relatively few work aiming to tackle the graph pre-training and transferring problems \cite{hu2020strategies,ying2018transfer}, given the fact that it faces several challenges such as designing graph pre-training strategies and mitigating negative transfer.
A GNN-based strategy proposed in\cite{hu2020strategies} combines both node- and graph-level pre-training to handle out-of-distribution samples when transferring. 
UDA-GCN developed by \cite{wu2020unsupervised} jointly integrates local and global consistency and facilitates knowledge transferring across graphs. 
GPT-GNN \cite{hu2020gpt} is another generative GNN framework that uses attribute/edge generations for pre-training on large-scale graphs.

In this work, \M~has largely expanded the training data including labeled, unlabeled, and augmented cascade graphs from different data domains, enabling the model to learn general and robust graph representations from abundant data, and then transferring the learned knowledge to downstream prediction tasks, at where the task-specific labeled data are scarce. 
\M~does not rely on special pretext tasks nor domain expertise and can be easily extended to other cascade graph applications. 

\section{Preliminaries}
\label{sec:preliminaries}

We now introduce the basic settings and describe the necessary preliminaries.
Information cascade is \textit{a behavior of information adoption by people}~\cite{guille2013information}, in various applications such as social networks of Twitter and Weibo, academic networks of authors and papers. The sequence of adoptions over time forms information cascade. If we have the diffusion path of each adoption, then the cascade graph \cite{cheng2014can,chen2019information} can be constructed, formally defined as:

\begin{definition}{\textbf{Information Cascade Graph}.} 
\textit{Given an information item $I$, e.g., a tweet or a paper, published at time $t_0$, over a period of time, item $I$ receives several adoptions, e.g., $M$ retweets or citations, then the sequence of adoptions composes an information cascade $C_I(t)$$ = \{(u_j, t_j)|j\in [1, M], t_j < t\}$, where user $u_j$ adopts the item $I$ at time $t_j$. Then a cascade graph can be defined as a diffusion tree $\mathcal{G}(t) = \{\mathcal{V}, \mathcal{E}\}$, where $\mathcal{V}$ denotes the set of users in $C_I(t)$ and $\mathcal{E}$ denotes the adoption relationship between users, e.g., retweeting or citing.}
\end{definition}

In this paper, we focus on the temporal-structural modeling of information cascades, i.e., given a sequence of cascade graphs, we aim to learn effective representations of cascade graphs for downstream cascade applications, e.g., outbreak prediction, recommendation, rumor detection, and user activation prediction. Most of this study focuses on \textit{information cascade popularity prediction} \cite{zhou2021survey}:

\begin{definition}{\textbf{Information Cascade Popularity Prediction}.}\label{def:popularity}
\textit{
Given an observed cascade graph $\mathcal{G}_i(t_o)$ at observation time $t_o$, the popularity prediction problem aims to predict the future popularity (or size) $P_i(t_p)$ of this cascade (graph) at a prediction time $t_p \gg t_o$. 
}
\end{definition}

Additional experiments conducted on another downstream task \textit{outbreak prediction} can be found in Section~\ref{subsec:transfer}. 
Other modalities of information cascades, e.g., global structures, individual user/item features, and content of texts/images, are not studied and left for future work.

\begin{table}[t]
    \small
    \caption{Mathematical symbols used in this paper.}
    \label{tab:symbol}
    \begin{tabular}{lp{6.9cm}}
    \toprule
        \textbf{Symbol} & \textbf{Description}  \\ \midrule
        $a_j^i$ & Attractiveness of node $u_j^i$ in cascade $C_i$. \\
        $B$ & Batch size. \\
        $\mathcal{V}_i, \mathcal{E}_i$ & Node and edge sets in cascade graph $\mathcal{G}_i$. \\
        $\mathcal{G}_i, \tilde{\mathcal{G}}_i$ & Cascade graph and augmented cascade graph. \\
        $\mathbf{h}_i$ & Latent representation used for downstream tasks. \\
        $M$ & Number of nodes in cascade graph. \\
        $N$ & Number of labeled cascades. \\
        $\mathcal{N}(u_j^i)$ & Neighboring nodes of $u_j^i$. \\
        $P_i(t_p)$ & Popularity of cascade $C_i$ at time $t_p$. \\
        $r_j^i$ & Remove probability of node $u_j^i$ in cascade $C_i$. \\
        $t_j^i$ & Adoption time of user $u_j^i$. \\ 
        $u_j^i$ & User in cascade $C_i$. \\ 
        $U$ & Number of unlabeled cascades. \\
        $\mathbf{z}_i$ & Latent representation used in contrastive loss. \\
    \bottomrule
    \end{tabular}
    
\end{table}

In order to explore and understand which parts of an unsupervised learning framework can benefit the learning of cascade representations, we concentrate on answering the following three questions. 

\vpara{Q1: Will unlabeled data improve the learning of cascade graph representation and prediction?}
Most of the supervised models are not possible to benefit from unlabeled data. In the context of graph-based cascade understanding, unlabeled graphs can be obtained at their early evolving stage, which is not able to make meaningful predictions. In traditional prediction models, these graphs are simply filtered out from training and evaluation \cite{cao2017deephawkes,zhao2015seismic,zhou2020variational}. 
Self-supervised approaches emphasize the importance of learning effective representations from a large amount of unlabeled data in an unsupervised and task-agnostic way. Unlabeled data can be easily obtained from one \cite{chen2020simple} or multiple datasets \cite{qiu2020gcc} for learning generic representations. Therefore, we believe that we can improve the cascade graph learning by incorporating unlabeled cascades in \M.

\vpara{Q2: Will data augmentation improve the cascade prediction and if so, how can we design augmentation strategies for cascade graphs?}
Since graph augmentation procedures have no analogues to texts or images due to the non-Euclidean, complex structures of graphs, how to design new strategies for graph data to improve model generalization ability and to benefit downstream prediction tasks becomes critical both for graph and contrastive learning. 
As we discussed earlier, previous graph data augmentation techniques mainly focus on graph neural networks and node/graph classifications \cite{sun2020infograph,rong2019dropedge,chen2020measuring}, and most of them only consider edge and feature manipulations while ignore node handling. This calls for an urgent research to devise \textit{cascade graph-specific} augmentation strategies \cite{zhao2020data}. In \M, we propose a novel strategy for cascade graph data augmentation: 
we simulate the mechanism of information diffusion in social networks (or other similar networks, e.g., news and academics), where we first traverse every node in the graph in their adoption time order, and then for each node we compute its attractiveness probability. Depends on their degrees in the cascade graph, current nodes can attract new adopters or lose existing followers. 

\vpara{Q3: Will contrastive self-supervised learning framework improve the cascade learning and prediction?}
Pre-training models has recently received a great deal of attention and achieved good performance in both linguistic and visual tasks. However, their applicability on information cascades is still underexplored by research community, to the best of our knowledge. In this study, based on the idea of pre-training \cite{chen2020big,sun2020infograph,velickovic2019deep}, we develop \M~where the selection of cascade graph encoder network is generic. Any graph representation learning models and graph neural networks, or other specifically designed cascade learning models (e.g., DeepCas \cite{li2017deepcas}, VaCas \cite{zhou2020variational}, Coupled-GNNs \cite{cao2020popularity}) can be used as cascade graph encoders. 
We implement and compare these techniques in \M, aiming at shedding some light on the ability of unsupervised cascade representation learning and seeking additional performance improvement for downstream cascade prediction tasks.

We finish this section with a summary of the symbols that will be used in the rest of the paper, presented in Table~\ref{tab:symbol}.

\begin{figure}[t]
    \centering
    \includegraphics[width=\linewidth]{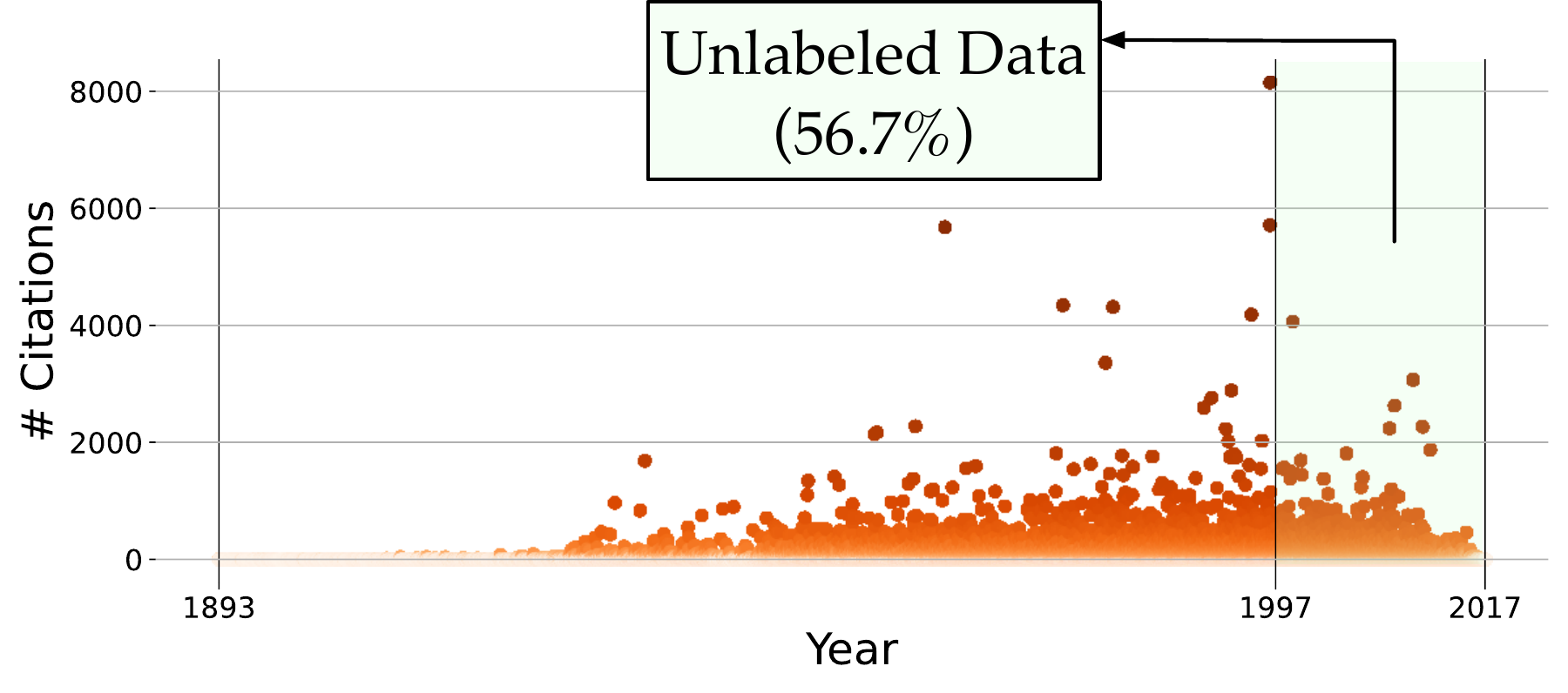}

    \caption{All 616,316 papers in the APS dataset spanning from Jul, 1893 to Dec, 2017. Papers published after Dec, 1997 have no appropriate labels if we want to predict their citation numbers after 20 years of publication.}
    \label{fig:aps-year-citations}
    
\end{figure}

\section{CCGL Framework}
\label{sec:methodology}
\begin{figure*}[t]
    \centering
    \includegraphics[width=\textwidth]{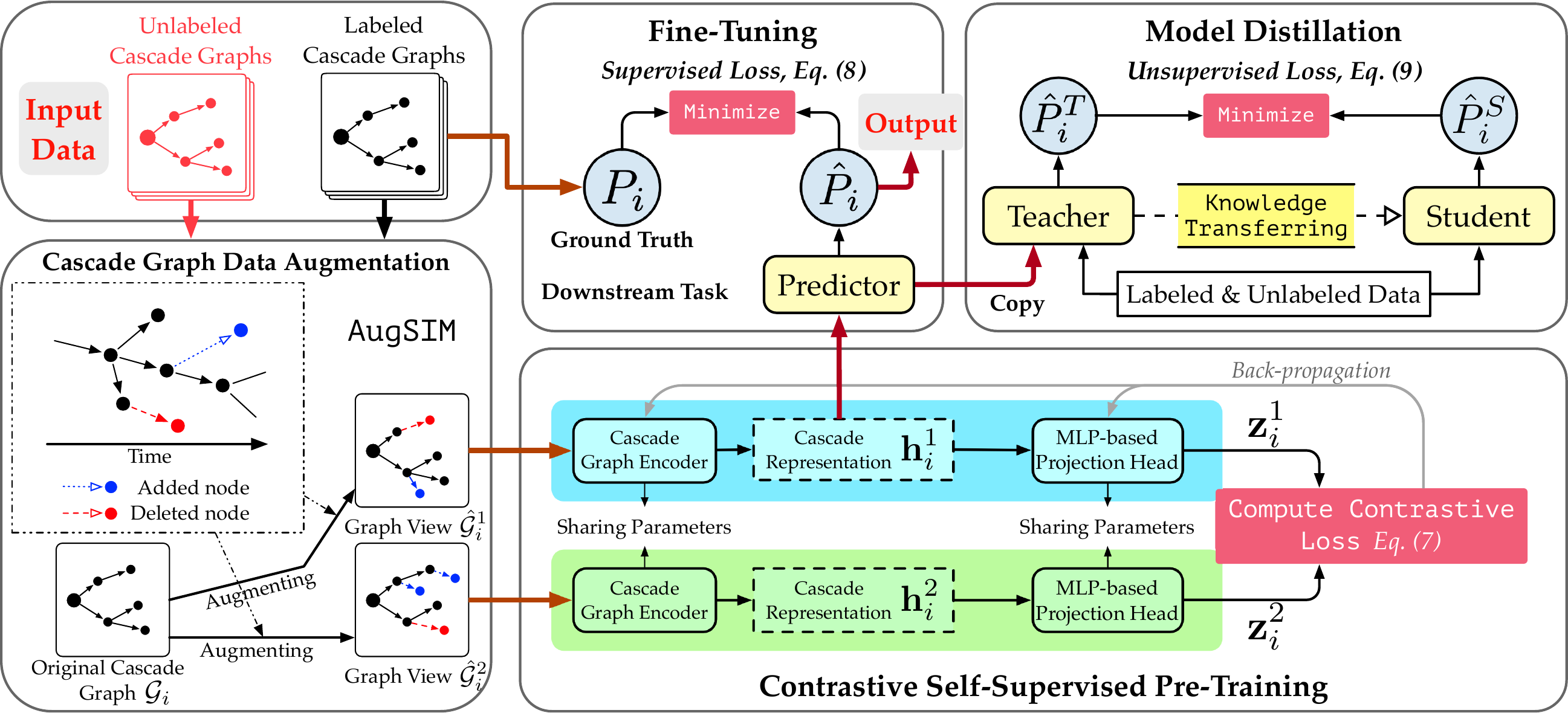}
    
    \caption{The overview of our proposed \M~framework for learning cascade graph representations. It consists of three components: (i) data augmentation strategy AugSIM designed for information cascade graphs; (ii) unsupervised pre-training the \M~framework by minimizing the contrastive losses on cascade graph representations in a \textit{task-agnostic} way; and (iii) fine-tuning and further distilling the \M~framework in a \textit{task-specific} way. Pre-training and distillation stages utilize both labeled and unlabeled data. Predictor and Teacher are the same network.}
    \label{fig:framework}
\end{figure*}

We now discuss the main aspects of \M~sketched in \figurename~\ref{fig:framework}, which consists of three major components: 

\noindent(i) Cascade graph data augmentation. To learn a more generic and transferable knowledge of graph cascade representations, we design an augmenting strategy using both labeled and unlabeled data to capture some variation and uncertainty via an information diffusion process simulation. See details in Section \ref{subsec:4.2}. \\
\noindent(ii) Self-supervised pre-training. We leverage the contrastive pre-training framework to learn abstract-level representations when encoding cascade graphs, which can help alleviate overfitting that typically occurs in many generative representation learning paradigms where fine-grained feature-level representations are learned (linking data to a specific task). See details in Section \ref{subsec:4.3}.\\
\noindent(iii) Model fine-tuning and distillation. For certain downstream tasks, we fine-tune the pre-trained model using labeled data. We also distill the model to make it more generalized and robust. Without having the distillation, we usually end up with a task-specific model where negative transferring happens when we apply the learned model to different datasets or other tasks. See details in Section \ref{subsec:4.4}.

\subsection{Utilizing Unlabeled Cascade Graphs}\label{subsec:4.1}

In typical information cascade prediction tasks, 
unlabeled data are simply excluded from training and evaluation. For example, when predicting the popularity of an information item, say, a tweet or a scientific paper, we first need to observe its early growth trend and try to predict its future popularity at a given time. In \cite{cao2017deephawkes}, given a scientific paper and its observed citations in the first few years, the authors predict its citations in 20 years. However, the papers with life span less than 20 years, i.e., without appropriate labels specified, are simply filtered out from the dataset. As shown in \figurename~\ref{fig:aps-year-citations}, for papers in the APS dataset, only about 43.3\% papers have been used during training and evaluation, leaving others as unlabeled cascades.
That is to say, this prediction system is only applicable to papers published at least 20 years ago, since only older papers (those with labels) are trained and evaluated. 

If we use this system to predict citations of newly published papers, we are biased in favor of past publications, and ignore recent diffusion characteristics during paper dissemination. 
To address such issues, we need to take unlabeled cascades into consideration, and learn representations of both labeled and unlabeled cascades in an unsupervised manner, which we detail next. 

\subsection{Augmenting Information Cascade Graphs}\label{subsec:4.2}

Existing data augmentation strategies designed in self-supervised learning frameworks are specific for learning language or visual representations, and basically built based on GNNs \cite{chen2020simple,he2020momentum,qiu2020gcc}. They are not directly applicable for cascade learning, due to the following reasons: 
(i) there are no straightforward ways to project text/image augmentation strategies to graphs; 
(ii) a tree-structured cascade graph starts from a root node (e.g., a newly published post or paper), and then diffuses and dynamically evolves to larger audience (e.g., retweets or citations). Adding/deleting nodes or edges in an arbitrary way may substantially change the structure of a cascade graph (e.g., deleting any edge would result in a disconnected $\mathcal{G}_i$); and (iii) nodes in a cascade graph are temporally characterized, i.e., the adoption time of nodes is of paramount importance to model cascade graphs, since cascade graphs with similar number of nodes, or even with the same structures, may have very different temporal behaviors \cite{zhou2021survey,cheng2014can}.  To address the above three challenges, we propose a novel and effective cascade graph data augmentation procedure: {\verb|AugSIM|}.

\vpara{AugSIM: \textit{\underline{Aug}menting cascade graphs by \underline{SIM}ulating an information diffusion process}.} 
In order to create different graph views of a target cascade graph for the subsequent contrastive prediction tasks, while also capturing a certain degree of similarity of graph topology or node/edge features, we propose a simple but effective augmentation procedure based on user's influence and adoption time. 
For each user $u_j^i$ in a cascade graph $\mathcal{G}_i$, we compute an \textit{attractiveness probability} $a_j^i$ to manage the node adding process:
\begin{equation}\label{equ:attractiveness}
    a_j^i = \eta_i \frac{\text{degree}(u_j^i)}{\sum_{u_k^i \in \mathcal{V}} \text{degree}(u_k^i)},
\end{equation}
where the \textit{augmentation strength} $\eta_i$ denotes a cascade-level hyper-parameter controlling the number of added nodes to $\mathcal{G}_i$. The added node $u_{\text{new}}^i$ connects to $u_j^i$ and should be assigned with an adoption time $t_{\text{new}}^i \in [t_j^i, t_o]$ as a node feature. The adoption time $t_j^i$ of an item can be viewed as an instance of human reaction time \cite{crane2008robust}. We compute both a local (cascade-level) adoption time $t^i_{\text{local}}$ and a global (dataset-level) adoption time $t_{\text{global}}$ to specify the adoption time of the newly added node:
\begin{equation}\label{equ:adoption-time}
    t_{\text{new}}^i = t_j^i + \theta_t t_{\text{local}}^i + (1 - \theta_t) t_{\text{global}},
\end{equation}
where $\theta_t$ is a weight parameter balancing two adoption times, $t_{\text{local}}^i$ is the average adoption time $(\frac{1}{|\mathcal{V}_i|}\sum_{j\in |\mathcal{V}_i|} t_j^i)$ of cascade $C_i$, and $t_{\text{global}}$ is the global adoption time drawn from an exponential distribution:
\begin{equation}\label{equ:kernel}
    f(t;\lambda) \sim \lambda e^{-\lambda t}, \quad\text{with}\quad \lambda > 0. 
\end{equation}
In Eq.~\eqref{equ:kernel} above, $\lambda$ is the rate parameter which can be fitted empirically from all adoption times in the dataset. 
After traversing every node in the cascade graph, we add a set of nodes $\mathcal{V}_{i}^{\text{new}} = \{u_{\text{new}, k}^i | k=1, 2, \dots\}$, node features $\{t_{\text{new}, k}^i | k=1, 2, \dots\}$, and edges $\mathcal{E}_i^{\text{new}}$, into $\mathcal{G}_i$ where each node is connected to its parent node. The expected number of added nodes in $\mathcal{V}_i^{\text{new}}$ is controlled by $\eta_i$, i.e., $E(|\mathcal{V}_{i}^{\text{new}}|) = \sum_{j=1}^{|\mathcal{V}_i|} a_j^i = \eta_i$. Correspondingly, the number of added edges is the same as the number of added nodes. 
We note that the added nodes are new to the cascade graph. 
Next we discuss how to remove nodes/edges in a cascade graph. Similarly, we traverse every leaf node in the expanded node set $\mathcal{V}_i \cup \mathcal{V}_{i}^{\text{new}}$. For each leaf node $u_j^i \in \mathcal{V}_i^{\text{leaf}}$, we compute a \textit{removal probability} $r_j^i$, defined as:
\begin{equation}\label{equ:remove}
    r_j^i = \eta_i \frac{\text{degree}(\text{parent}(u_j^i))}{\sum_{u_k^i \in \mathcal{V}_i^{\text{leaf}}} \text{degree}(\text{parent}(u_k^i))},
\end{equation}
where $v_j^i$ is the parent node of $u_j^i \in \mathcal{V}_i^{\text{leaf}}$. 
The expected number of removed nodes/edges is $\sum_{j=1}^{\mathcal{V}_i^{\text{leaf}}} r_j^i = \eta_i$. 

For simplicity, the node removal process is only conducted on leaf nodes, as the main cascade graph structure is maintained. However, other more sophisticated strategies can be used to simulate the information diffusion, such as: 
(i) allow added nodes to attract more followers; 
(ii) remove not only leaf nodes but also their parents; 
(iii) consider more features, such as the number of followers/followees, the number of citations or $h$-index of authors, as a surrogate of user influence to choose (added/removed) nodes or to specify appropriate adoption times; 
(iv) adopt stochastic point processes such as Poisson process and Hawkes self-exciting process, which are frequently used to describe the cascading behaviors in information diffusion \cite{crane2008robust,zhao2015seismic}, and use them as generative models to augment cascade graphs or expand training data \cite{hu2020gpt}. 
Potential improvement of cascade graph data augmentation is subject to future work. Here we use node degrees and adoption times to augment cascade graphs, creating different but similar views of graphs for later contrastive modeling. 
\verb|AugSIM| strategy can be viewed as an instance of re-diffusion of an information in a network, which preserves the basic patterns of diffusion and introduces some variation and uncertainty.

We also designed another two augmentation strategies for comparison: \texttt{AugRWR} and \texttt{AugAttr}, cf. Section~\ref{sub:augmentation}.

\subsection{Learning Cascade Graph Representation via Contrastive Self-Supervised Learning}\label{subsec:4.3}

With data augmentation for cascade graph in place, we now introduce the \M~framework for learning generalized cascade representations without label supervision. 

\vpara{Data augmentation.} 
We first use one of the augmentation strategies to create related views of the same cascade graph. Given $\mathcal{G}_i$, we augment this graph twice to create two different but similar views, denoted as $\tilde{\mathcal{G}}_i^1$ and $\tilde{\mathcal{G}}_i^2$. These two augmented graphs are considered as a positive pair $(\tilde{\mathcal{G}}_i^1, \tilde{\mathcal{G}}_i^2)$ in the subsequent contrastive learning. 

\vpara{Cascade graph encoding.} 
We then encode to represent a cascade graph to a vector while capturing temporal and structural information in the graph. The choice of cascade graph encoder is not confined to a particular approach. Any encoder that can map the sparse cascade graph into a dense representation vector is qualified. We employ a graph encoder from a state-of-the-art cascade prediction model VaCas \cite{zhou2020variational}, which has two main components: (i) a graph embedding based on spectral graph wavelets;
and (ii) a bi-directional GRU-based network to learn contextualized user behaviors in cascade data. Note that this model is equivalent to Cas-RNN as described in \cite{zhou2020variational}.
These two components map the cascade graph $\mathcal{G}_i$ to a fixed-length representation $\mathbf{h}_i \in \mathbb{R}^{d_{\mathbf{h}}}$. To further understand the relations among latent factors and avoid possible noises in representation, we follow a prior study~\cite{chen2020simple} and add a MLP-based projection head to project $\mathbf{h}_i$ to a new representation $\mathbf{z}_i \in \mathbb{R}^{d_{\mathbf{z}}}$. This has been demonstrated a significant improvement for some applications \cite{chen2020big,chen2020improved}, as well as ours. We experiment with different designs of projection head (cf. Section~\ref{subsec:results}).
\begin{align}\label{equ:encoder}
    \mathbf{h}_i^1 &= \text{cascade\_encoder}(\tilde{\mathcal{G}}_i^1), & \mathbf{z}_i^1 = \text{MLP}(\mathbf{h}_i^1),\\
    \mathbf{h}_i^2 &= \text{cascade\_encoder}(\tilde{\mathcal{G}}_i^2), & \mathbf{z}_i^2 = \text{MLP}(\mathbf{h}_i^2). 
\end{align}
Note that the projection head only takes part in an unsupervised learning stage, i.e., we still use $\mathbf{h}_i$ for subsequent downstream task fine-tuning, and use $\mathbf{z}_i$ for computing the contrastive loss and optimizing the \M~framework. 

\vpara{Contrastive loss.} 
In order to train our \M~framework, following \cite{chen2020simple}, the contrastive learning loss is defined to maximize the latent similarity between two augmented views of the same cascade graph, and discriminate between a positive pair $(\tilde{\mathcal{G}}_i^1, \tilde{\mathcal{G}}_i^2)$ and all other negative pairs in a mini-batch. 
Specifically, \M~first randomly samples $B$ cascade graphs, and then augments each graph twice to obtain $2B$ augmented cascade graphs. For a pair of positive graphs in a mini-batch, we leave the remaining $2B-2$ cascade graphs as negative samples. Given a similarity function $\text{sim}(\cdot, \cdot)$ over two vectors (in our case the cosine similarity), the contrastive loss function for a positive pair $(\tilde{\mathcal{G}}_i^1, \tilde{\mathcal{G}}_i^2)$ is defined as:
\begin{equation}\label{equ:contrastive-loss}
    \mathcal{L}_{1, 2}^{\text{contrastive}} = -\log \frac{\exp{(\text{sim}(\mathbf{z}_i^1, \mathbf{z}_i^2)/\tau)}}{\sum_{k=1}^{2B} \mathds{1}(k\neq i)\exp{(\text{sim}(\mathbf{z}_i^1, \mathbf{z}_i^k)/\tau)))}},
\end{equation}

\noindent where $\mathds{1}(\cdot)$ is an indicator function, $\tau$ is a temperature parameter. 
This contrastive loss function for unsupervised learning is noted as InfoNCE \cite{oord2018representation} or NT-Xent \cite{chen2020simple}, that has been widely used in previous SSL models \cite{he2020momentum,qiu2020gcc,wu2018unsupervised}. During pre-training, positive samples (two new views drawn from the original cascade graph) are attracted together in the representation space, while negative samples are repelled away from positive samples. In this way, we learn cascade graph representations in an abstract way, without linking them to any specific downstream task labels/signals.

\vpara{Discussion of loss mechanism.} 
Since contrastive learning often requires a large number of negative samples to contrast, maintaining larger mini-batches/dictionaries and building larger network architectures are non-economic and limited by computational resources, although improve the model performance. For example, SimCLR \cite{chen2020simple} uses a batch size as large as 8,192 (with 16,382 negative samples per positive pair) in a platform of 128 TPU v3 cores. The learning ability of end-to-end models is constrained by batch sizes. Some of them rely on special pretext tasks, which may change the network architecture, such as constraining the receptive field size \cite{bachman2019learning,hjelm2018learning}, patching the graph into sub-structures \cite{sun2020infograph}, and limiting their generalization performance. 
Memory bank \cite{wu2018unsupervised} and momentum update \cite{he2020momentum} are another line of contrastive mechanisms that decouple the number of negative samples from the mini-batch size, while providing smooth and consistent encoder update. 
Previous works have discussed the impact of end-to-end and memory mechanisms. For example, \cite{he2020momentum} found that end-to-end models have competitive performance when the batch size is small. \cite{qiu2020gcc} and \cite{chen2020big} concluded that memory mechanism only provides marginal improvement when a large batch size is used. However, larger mini-batch requires more GPU/TPU memory and longer training time. 

\subsection{Fine Tuning and Distilling \M~for Downstream Cascade Prediction Tasks}\label{subsec:4.4}

A general model is obtained via the contrastive learning by far, but requires a further fine-tuning using labeled data for certain downstream tasks. In this paper, we focus on the information cascade popularity prediction (cf.~Definition~\ref{def:popularity}) as our primary downstream task. 

\vpara{Fine-tuning.} 
After unsupervised training of \M~with unlabeled and augmented cascade graphs in a \textit{task-agnostic} way, we use labeled cascade graphs to fine-tune the \M~framework in a \textit{task-specific} way. 
Following \cite{chen2020big}, the MLP-based projection head used for contrastive learning can be fully discarded (i.e., only the cascade graph encoder is used for fine-tuning), partially discarded (i.e., an encoder followed by fully connected layers for fine-tuning), or fully included (i.e., we use $\mathbf{z}_i$ for both downstream fine-tuning task and contrastive learning). Detailed discussion of projection head is provided in Section~\ref{subsec:results}. 
For $N$ observed training cascade graphs, the training loss is defined as mean logarithmic squared error (MSLE): 
\begin{equation}\label{equ:sup-loss}
    \mathcal{L}^{\text{supervised}} = \frac{1}{N} \sum_{i=1}^N \left( \log \hat{P}_i(t_p) - \log P_i(t_p) \right)^2,
\end{equation}
where $P_i(t_p), \hat{P}_i(t_p)$ are the true and the predicted popularity, respectively. 

\vpara{Semi-supervised learning and model distillation.} 
Following previous studies \cite{chen2020simple,sun2020infograph}, we use the unsupervised contrastive loss Eq.~\eqref{equ:contrastive-loss} for pre-training and supervised loss Eq.~\eqref{equ:sup-loss} for fine-tuning. Unsupervised contrastive loss $\mathcal{L}^\text{contrastive}$ is computed by all positive pairs, forcing the model to discriminate augmented views for labeled and unlabeled cascade graphs. Supervised loss $\mathcal{L}^\text{supervised}$ is computed by labeled data for learning to predict future popularity. 
However, this loss combination setting may suffer from ``negative transfer'' \cite{hu2020strategies}. 
Inspired by \cite{chen2020big,sun2020infograph}, we adopt two separate networks to mitigate this issue -- a teacher network copied from the fine-tuned predictor and a student network started from the scratch. We enforce the prediction of the student network as similar as possible to the teacher network by minimizing the following revised loss function:
\begin{align}\label{equ:teacher-student}
    \mathcal{L}^{\text{semi}} &= \frac{1}{N+U} \sum_{i=1}^{N+U}  \left(\log \hat{P}_i^T(t_p) - \log \hat{P}_i^S(t_p) \right)^2  ,
\end{align}
where $N$ is the number of labeled samples and $U$ is the number of unlabeled samples, $\hat{P}_i^T(t_p)$ and $\hat{P}_i^S(t_p)$ are predictions of the teacher and student networks, respectively.
The weights of the teacher network are fixed and the weights of the student network are updated under Eq.~\eqref{equ:teacher-student}.
In this way, \M~benefits from both labeled and unlabeled data: the teacher network produces pseudo labels to the student network. The architecture of the student network can be identical to the teacher network (self-distillation), or be a smaller network to distill.

\subsection{Connection to Mutual Information Maximization}\label{sub:mim}
In theory, learning information cascade graph representation is to maximize the mutual information (MIM) \cite{oord2018representation,bachman2019learning} $\mathcal{I}_{\text{MI}}(\Phi(\hat{\mathcal{G}}_i^1);\Phi(\hat{\mathcal{G}}_i^2))$ between two augmented cascade graph views $\hat{\mathcal{G}}_i^1$ and $\hat{\mathcal{G}}_i^2$ (which should be classified in the same category or located closer in the embedding space) by a neural network $\Phi(\cdot)$. The network consists of a graph encoder maps the cascade graph $\mathcal{G}_i$ to $\mathbf{h}_i$ for downstream tasks and a MLP-based projection head maps the $\mathbf{h}_i$ to $\mathbf{z}_i$ for contrastive learning. 
From the perspective of probability, given two random variables $\mathbf{z}_i^1$ and $\mathbf{z}_i^2$, the model is forced to discriminate between positive pairs from joint distribution $p(\mathbf{z}_i^1 ,\mathbf{z}_i^2)$ and negative pairs from the product of marginals $p(\mathbf{z}_i^1)p(\mathbf{z}_i^2)$. Minimizing the loss \eqref{equ:contrastive-loss} is equivalent to maximize a lower bound on the mutual information $\mathcal{I}_{\text{MI}}(\mathbf{z}_i^1; \mathbf{z}_i^2)$ as:
\begin{align}\label{equ:mim}
    \mathcal{I}_{\text{MI}}(\mathbf{z}_i^1; \mathbf{z}_i^2)
    &= \mathbb{E}_{p(\mathbf{z}_i^1, \mathbf{z}_i^2)}\left[ \log \frac{p(\mathbf{z}_i^1, \mathbf{z}_i^2)}{p(\mathbf{z}_i^1)p(\mathbf{z}_i^2)} \right] \\ 
    &\geq \log(2B) - \mathbb{E}\left[ \mathcal{L}_{1, 2}^{\text{contrastive}}(\mathbf{z}_i^1, \mathbf{z}_i^2, 2B) \right]. 
\end{align}
The key factor of contrastive frameworks is the design of different data views. In \M, we propose \verb|AugSIM| to augment cascade graphs by simulating a new information diffusion process based on existing observation. \verb|AugSIM| is able to preserve the high-level shared information between graph views while also capturing the variation and uncertainty during diffusion, i.e., the learned cascade graph representations should resist to random perturbations. On one hand, the model intends to discriminate augmented positive/negative graph views as much as possible (for maximizing the mutual information). On the other hand, the model is optimized to ignore trivial differences in graph views to retain only the necessary information (for minimizing the prediction error of downstream tasks).

\begin{table}[t]
    \footnotesize
    \centering
    \caption{Run time analysis on 58,489 Weibo retweet cascades. }
    \label{tab:run-time}
    \begin{tabular}{lrr}
    \toprule
        \textbf{Stage}  & \textbf{Supervised} & \textbf{\M} \\ \midrule
        Augmentation    & - & 71.5s (1.2ms per) \\ 
        Pre-processing  & 17.5m (27ms per) & 53.5m (55ms per) \\
        Pre-training    & - & 14.0m (14ms per) \\
        Training/Fine-tuning        & 23.8m (36ms per) & 14.9m (23ms per) \\
        Distillation    & - & 36.5m (28ms per) \\
        Total run time  & 41.3m (63ms per) & 107.4m (110ms per) \\ 
    \bottomrule
    \end{tabular}
\end{table}

\subsection{Computational Complexity}\label{subsec:4.5}

Compared to supervised baselines, \M~framework has three more components that bring extra computational overhead: (i) graph data augmentation; (ii) contrastive self-supervised pre-training; and (iii) model distillation. 

\vpara{Graph data augmentation.} Let $|\mathcal{V}|$ and $|\mathcal{V}^{\text{leaf}}|$ be the number of nodes and number of leaf nodes (after adding process) in a cascade graph, augmentation strategy \verb|AugSIM| needs to traverse every node in graph to add nodes and then traverse every leaf node in a graph to remove nodes, where the time complexity is $\mathcal{O}(|\mathcal{V}| + |\mathcal{V}^{\text{leaf}}|)$, approximately to or less than $\mathcal{O}(2|\mathcal{V}|)$, which is linear to the number of nodes in a graph. 
For 58K Weibo cascades, \verb|AugSIM| only takes $\sim$71.5s (or $\sim$1.3ms per cascade).

\vpara{Pre-training, fine-tuning and distillation.} The number of trainable parameters in pre-training network is on par with supervised baselines (708K vs. 686K). Extra parameters are mainly from the projection head (depending on depth and width of the head). 
\M~costs $\sim$27s for one epoch of pre-training, $\sim$13s for one epoch of fine-tuning, and $\sim$33s for one epoch of distillation. 
The experiments were conducted on Ubuntu 20.04, with 48GB RAM, an Intel\textregistered~Core\texttrademark~i7-8700K CPU, single NVIDIA 1080Ti GPU. 
\M~is implemented by Python 3.7, TensorFlow 2.3 with CUDA 10.1. 
The run time analysis with comparison to supervised models is summarized in Table~\ref{tab:run-time}. We use default hyper-parameter settings (cf. Table~\ref{tab:hyper-parameters}).

\section{Experiments}
\label{sec:experiments}
\begin{table}[t]
    \footnotesize
    \centering
    \setlength{\tabcolsep}{4.7pt}
    \caption{Descriptive statistics of five information cascade datasets.}
    \label{tab:datasets}
    \begin{tabular}{@{}lrrrrr@{}}
    \toprule
        \textbf{Statistic}      & \textbf{Weibo}    & \textbf{Twitter}  & \textbf{ACM}   & \textbf{APS}  & \textbf{DBLP} \\ \midrule
        \# labeled cascades     & 39,076    & 18,198    & 12,988    & 27,802    & 4,879 \\
        \# unlabeled cascades   & 19,413    & 7,396     & 20,013   & 44,921    & 264 \\
        Avg. \# nodes           & 174.02    & 141.58    & 17.69     & 30.82     & 14.99 \\
        Avg. observed \# nodes  & 98.91     & 82.46     & 12.20     & 17.54     & 10.80 \\
        Avg. path length        & 2.2462    & 2.2111    & 2.0805    & 2.9164    & 2.5413 \\
    \bottomrule
    \end{tabular}
\end{table}

\begin{table}[t]
    \footnotesize
    \centering
    \caption{Hyper-parameter settings.}
    \label{tab:hyper-parameters}
    \begin{tabular}{@{}llr@{}}
    \toprule
        \textbf{Parameter}    & \textbf{Search Space} & \textbf{Value} \\ \midrule
        Augmentation strength $\eta$ & $\{0.05, 0.1, 0.2, 0.5, 1.0\}$ & 0.1 \\
        Augmentation strategy & {\verb|AugSIM|},  {\verb|AugRWR|}, \texttt{AugAttr} & {\verb|AugSIM|} \\
        Batch size $B$ & $\{16, 32, 64, 256\}$ & 64 \\
        Early stopping patience & - & 20 \\
        Embedding dimension & $\{16, 32, 64, 128, 256, 512\}$ & 64 \\
        Learning rate & - & $5e^{-4}$ \\
        Pre-training epochs & $\{10, 20, 30, 50, 100, 200, 500\}$ & 30 \\
        Projection head (100\%) & From 0 up to 4 layers & $4$-$1$ \\ 
        Projection head (10\%) & From 0 up to 4 layers & $4$-$4$ \\ 
        Projection head (1\%)  & From 0 up to 4 layers & $4$-$3$ \\ 
        Restart probability $s$ & - & 0.2 \\
        Model size & $\{1\times, 2\times, 4\times, 8\times, 16\times\}$ & $4\times$ \\
        RWR walking steps $\gamma$ & $\{1.0, 2.0, 3.0, 5.0\}$ & 3.0 \\
        Temperature $\tau$ & $\{0.05, 0.1, 0.2, 0.5, 1.0, 2.0\}$ & 0.1 \\
    \bottomrule
    \end{tabular}
\end{table}

Following common protocols of unsupervised and semi-supervised learning  \cite{he2020momentum,chen2020big,qiu2020gcc}, 
we summarize experimental settings and several social/scientific cascade datasets in Section~\ref{subsec:settings} and \ref{subsec:datasets}, respectively. Baselines and their configurations are provided in Section~\ref{subsec:baselines}. 
We have the discussions of experimental results, several observations, and ablation studies in Section~\ref{subsec:results}.  
We conduct knowledge transferring experiments (among different tasks and datasets) in Section~\ref{subsec:transfer}. 

\subsection{Experimental Settings}\label{subsec:settings}

For all experiments of \M~and baselines, unless otherwise specified, we uniformly adopt the following settings for a fair comparison.
We use Adam optimizer, and each dataset is divided into training (50\%), validation (10\%), and test (40\%) sets, plus unlabeled data. The pre-training (fine-tuning) is early stopped when training (validation) loss has not declined with a patience of 20 epochs. We report MSLE with logarithmic base 2 following \cite{li2017deepcas,cao2017deephawkes,zhou2020variational}. Cascade graphs with nodes $|\mathcal{V}(t_o)|<10$ are filtered out, and for graphs with $|\mathcal{V}(t_o)|>100$, we select the first 100 nodes (sorted by adoption time). 

We manually tune the model by searching the hyper-parameter space. Table~\ref{tab:hyper-parameters} lists hyper-parameters, searching space, and their default values used throughout this paper.

\subsection{Datasets}\label{subsec:datasets}

We used five large-scale publicly available cascade datasets, which can be categorized into two types: \textit{social} and \textit{scientific}. Detailed statistics of datasets is presented in Table~\ref{tab:datasets}.

\noindent$\bullet$ \textbf{Weibo} \textit{retweet} cascade dataset is introduced by \cite{cao2017deephawkes}. For cascades we set the observation time $t_o$ to 1 hour, prediction time $t_p$ to 24 hours. 

\noindent$\bullet$ \textbf{Twitter} \textit{hashtag} cascade dataset is collected by \cite{weng2013virality}. The cascade graphs are built by adopting, retweeting, and mentioning relationships. We set observation time $t_o$ to 2 days, prediction time $t_p$ to 32 days. 

\noindent$\bullet$ \textbf{ACM} \textit{citation} cascade dataset is rearranged from Association for Computing Machinery \cite{tang2008arnetminer} (released at Jan 20, 2017), which contains 2,385,057 scientific papers in the field of computer science. We set the observation time $t_o$ to 3 years, prediction time $t_p$ to 10 years.

\noindent$\bullet$ \textbf{APS} \textit{citation} cascade dataset is rearranged from American Physical Society at \url{https://journals.aps.org/datasets} (accessed on Jan 17, 2019), containing 616,316 papers published by 17 APS journals. We set the observation time $t_o$ to 3 years, prediction time $t_p$ to 20 years.

\noindent$\bullet$ \textbf{DBLP} \textit{citation} cascade dataset is from DBLP citation network V9 \cite{tang2008arnetminer} (released at Jul 3, 2017), containing 3,680,006 scientific papers. We set the observation time $t_o$ to 5 years, prediction time $t_p$ to 20 years. We filter papers with less than 5 citations within the first 5 years, that is $|\mathcal{V}(t_o)| < 5$.

\begin{table*}[t]
    \centering
    \renewcommand{\arraystretch}{1.15}
    \caption{Prediction performance comparison of both supervised and semi-supervised baselines and our proposed \M~framework, under different experimental settings, measured by 10 runs of mean MSLEs (lower is better) on the Weibo dataset with varied label fractions.
    Results of another four datasets (Twitter, ACM, APS, DBLP) are in Table~\ref{tab:results-on-other-datasets}.
    We use a self-distilled student. 
    The bottom right number of MSLE is standard deviation, upper right is improvement compared to the Base at least {\color{orange}0.05} point. 
    Numbers in parentheses refer to equations. 
    }
    \label{tab:results}
    \begin{tabular}{lcllrrr}
    \toprule
        \textbf{Model} & \multicolumn{1}{l}{\textbf{Unlabel}} & \textbf{Augmentation} & \textbf{Loss mechanism} & \multicolumn{1}{r}{\textbf{Weibo (1\%)}} & \multicolumn{1}{r}{\textbf{Weibo (10\%)}} & \multicolumn{1}{r}{\textbf{Weibo (100\%)}} \\ \midrule
        \multicolumn{7}{l}{\textit{Supervised baselines:}} \\
        Feature & - & - & supervised \eqref{equ:sup-loss} & 4.94$_{\pm 0.70}$ & 3.35$_{\pm 0.15}$ & 3.01$_{\pm 0.17}$ \\
        DeepHawkes \cite{cao2017deephawkes} & - & - & supervised \eqref{equ:sup-loss}  & 4.39$_{\pm 0.10}$ & 3.41$_{\pm 0.03}$ & 3.24$_{\pm 0.02}$ \\
        node2vec \cite{grover2016node2vec}+BiGRU & - & - & supervised \eqref{equ:sup-loss} & 3.76$_{\pm 0.38}$ & 3.44$_{\pm 0.03}$ & 2.95$_{\pm 0.03}$ \\
        Base (rand. init.) \cite{zhou2020variational} & - & {\verb|AugSIM|} & supervised \eqref{equ:sup-loss} & 3.82$_{\pm 0.10}$ & 3.34$_{\pm 0.01}$ & 3.06$_{\pm 0.19}$ \\
        \rowcolor[gray]{.95} & - & - & supervised \eqref{equ:sup-loss}  & \textbf{3.58}$_{\pm 0.03}$ & \textbf{3.24}$_{\pm 0.09}$ & \textbf{2.77}$_{\pm 0.04}$ \\ \arrayrulecolor[gray]{.5} \midrule
        \multicolumn{7}{l}{\textit{Semi-supervised models fully fine-tuned or linear evaluated:}} \\
        AE & \ding{51} & - & reconstruction error & 4.00$_{\pm 0.38}$ & 3.66$_{\pm 0.57}$ & 2.78$_{\pm 0.04}$ \\
        VAE & \ding{51} & - & reconstruction error + ELBO & 3.96$_{\pm 0.44}$ & 3.63$_{\pm 0.71}$ & 2.78$_{\pm 0.09}$ \\
        \M~(ours, freeze) & \ding{51} & \verb|AugRWR| & supervised + contrastive \eqref{equ:contrastive-loss}+\eqref{equ:sup-loss} & 4.18$_{\pm 0.05}$ & 3.81$_{\pm 0.01}$ & 2.84$_{\pm 0.02}$ \\ 
        & \ding{55} & \verb|AugSIM| & supervised + contrastive \eqref{equ:contrastive-loss}+\eqref{equ:sup-loss} & 4.42$_{\pm 0.23}$ & 3.69$_{\pm 0.02}$ & 2.81$_{\pm 0.04}$ \\ 
        & \ding{51} & \verb|AugSIM| & supervised + contrastive \eqref{equ:contrastive-loss}+\eqref{equ:sup-loss} & 4.51$_{\pm 0.09}$ & 4.11$_{\pm 0.04}$ & 2.78$_{\pm 0.02}$ \\ 
        \M~(ours) & \ding{51} & \verb|AugAttr|              & supervised + contrastive \eqref{equ:contrastive-loss}+\eqref{equ:sup-loss} & 3.55$_{\pm 0.03}^{{\color{orange}+0.08}}$ & 3.07$_{\pm 0.09}^{{\color{orange}+0.17}}$ & 2.79$_{\pm 0.02}$ \\
        & \ding{51} & \verb|AugRWR|               & supervised + contrastive \eqref{equ:contrastive-loss}+\eqref{equ:sup-loss} & 3.49$_{\pm 0.05}^{{\color{orange}+0.09}}$ & 3.12$_{\pm 0.29}^{{\color{orange}+0.12}}$ & 2.74$_{\pm 0.02}$ \\
        & \ding{51} & \verb|AugSIM|     & supervised + contrastive \eqref{equ:contrastive-loss}+\eqref{equ:sup-loss} & 3.39$_{\pm 0.10}^{{\color{orange}+ 0.19}}$ & 2.94$_{\pm 0.03}^{{\color{orange}+ 0.30}}$ & 2.73$_{\pm 0.02}$ \\
        & \ding{55} & \verb|AugRWR|               & supervised + contrastive \eqref{equ:contrastive-loss}+\eqref{equ:sup-loss} & 3.44$_{\pm 0.05}^{{\color{orange}+0.14}}$ & 3.04$_{\pm 0.08}^{{\color{orange}+0.20}}$ & 2.75$_{\pm 0.04}$ \\
        & \ding{55} & \verb|AugSIM|               & supervised + contrastive \eqref{equ:contrastive-loss}+\eqref{equ:sup-loss} & 3.38$_{\pm 0.04}^{{\color{orange}+0.20}}$ & 2.95$_{\pm 0.04}^{{\color{orange}+0.29}}$ & 2.73$_{\pm 0.02}$ \\
        & \ding{51} & \verb|AugSIM|+\verb|AugRWR| & supervised + contrastive \eqref{equ:contrastive-loss}+\eqref{equ:sup-loss} & 3.35$_{\pm 0.08}^{{\color{orange}+0.23}}$ & 2.96$_{\pm 0.04}^{{\color{orange}+0.28}}$ & 2.71$_{\pm 0.02}^{{\color{orange}+0.06}}$ \\
        & \ding{51} & \verb|AugAttr|              & super.+unsup.+contr. \eqref{equ:contrastive-loss}+\eqref{equ:sup-loss}+\eqref{equ:teacher-student} & 3.50$_{\pm 0.05}^{{\color{orange}+0.08}}$ & 3.00$_{\pm 0.01}^{{\color{orange}+0.24}}$ & 2.77$_{\pm 0.01}$ \\
        & \ding{51} & \verb|AugRWR|               & super.+unsup.+contr. \eqref{equ:contrastive-loss}+\eqref{equ:sup-loss}+\eqref{equ:teacher-student} & 3.40$_{\pm 0.01}^{{\color{orange}+0.18}}$ & 3.02$_{\pm 0.01}^{{\color{orange}+0.22}}$ & 2.72$_{\pm 0.01}^{{\color{orange}+0.05}}$ \\ 
        \rowcolor[gray]{.95} & \ding{51} & \verb|AugSIM|               & super.+unsup.+contr. \eqref{equ:contrastive-loss}+\eqref{equ:sup-loss}+\eqref{equ:teacher-student} & \textbf{3.25}$_{\pm 0.02}^{{\color{orange}+0.33}}$& \textbf{2.86$_{\pm 0.01}^{{\color{orange}+0.38}}$}& \textbf{2.69}$_{\pm 0.00}^{{\color{orange}+0.08}}$ \\ 
        \rowcolor[gray]{.95} {\scriptsize(improves)} & & & & {\scriptsize9.2\%$\uparrow$} & {\scriptsize11.7\%$\uparrow$} & {\scriptsize2.9\%$\uparrow$} \\
    \arrayrulecolor[gray]{0} \bottomrule
    \end{tabular}
\end{table*}

\subsection{Baselines}\label{subsec:baselines}
To evaluate the performance of \M~and show the benefits of its three major components: a self-supervised task-agnostic contrastive model for generalizability, graph cascade augmentation to capture diffusion variation and uncertainty, and task-specific model fine-tuning and model distillation for knowledge transfer, we include several strong supervised and semi-supervised models as follows.

\subsubsection{Supervised models}
\noindent$\bullet$ \textbf{Feature}-based models extract hand-crafted features from information cascades to make predictions. Following \cite{cheng2014can,zhou2020variational}, we select several structural and temporal features: cumulative popularity series, time between root user and first adopter, mean time between the first and second half of adoptions, number of leaf nodes, mean of node degrees, mean and max length of diffusion paths; extracted features are 
then fed into fully connected layers for training and evaluation. 

\noindent$\bullet$ \textbf{node2vec} \cite{grover2016node2vec} is a node embedding technique based on random walk. We leverage it to obtain embeddings of nodes in cascade graph, and then feed them into Bi-GRUs followed by MLPs to make predictions. 
We set the embedding dimension to 128, window size to 10, walk length to 30, number of walks to 200, and both parameters $p$ and $q$ for neighborhood sampling to 1.
The implementation used is: \url{https://github.com/eliorc/node2vec}.

\noindent$\bullet$ \textbf{DeepHawkes} \cite{cao2017deephawkes} model predicts the popularity of information cascades by combining Hawkes process with deep learning. 
The embedding dimension is 64, learning rate is $5e^{-3}$, embedding learning rate is $5e^{-4}$, dropout prob. is $0.8$, and time interval is 5 minutes.
The implementation used is: \url{https://github.com/CaoQi92/DeepHawkes}.

\noindent$\bullet$ \textbf{Base} model~\cite{zhou2020variational}, a standard supervised benchmark, has the same network architecture (i.e., cascade graph encoder + MLPs) as our fine-tuned \M~but can not learn from unlabeled data. It lacks of contrastive pre-training and model distillation. 
For fairness, hyper-parameters in the Base model are exactly the same as \M. 

\subsubsection{Semi-supervised models}
\noindent$\bullet$ \textbf{Auto-encoder (AE)} and \textbf{variational auto-encoder (VAE)}
\cite{kingma2013auto} are deep generative models for unsupervised representations learning. 
The encoder and decoder of (V)AEs are both composed by GRUs and MLPs. The sequence of node embeddings in cascade graphs are first fed into the encoder to generate a dense latent representation $\mathbf{z}$, used by the decoder to reconstruct the input (i.e., node embeddings). The AE is optimized by minimizing the reconstruction loss; the objective for VAE includes an evidence lower bound (ELBO) term. 
We set the pre-training epoch to 100. 

\subsubsection{Other data augmentation strategies}\label{sub:augmentation}
To demonstrate the superiority of our \texttt{AugSIM}, we implement two additional augmentation strategies.

\noindent$\bullet$ \textbf{AugRWR}: \textit{\underline{Aug}menting cascade graphs by random walk with restart (\underline{RWR}) and sub-graph induction}, is partly inspired by GCC \cite{qiu2020gcc}. By repeating the RWR process, we collect a subset of nodes in $\mathcal{V}_i$, denoted as $\mathcal{V}_i^{\text{rwr}}$. Then the augmented graph $\tilde{\mathcal{G}}_i$ can be induced by removing the nodes in $\mathcal{V}_i \setminus \mathcal{V}_i^{\text{rwr}}$. We restrict the walking step at most $\gamma |\mathcal{V}_i|$ times, to avoid identical views for small-sized graphs and insufficient walk for large-sized graphs. The transition probability of RWR is specified by the degrees of neighboring nodes in graph: 
$p(v_j^i \in \mathcal{N}(u_j^i) | u_j^i) = \frac{\text{degree}(v_j^i)}{\sum_{v_k^i \in \mathcal{N}(u_j^i)} \text{degree}(v_k^i) },$
here $\mathcal{N}(u_j^i)$ is the set of neighboring nodes of $u_j^i$. 

\noindent$\bullet$ 
\textbf{AugAttr}: \textit{\underline{Aug}menting cascade graphs by replacing node \underline{Attr}ibutes}, is partly inspired by NodeAug \cite{wang2020nodeaug}. We alter node attribute (in our case the adoption time $t_j^i$) by iteratively and randomly sampling a new time $t_{j, \text{new}}^i$ during timespan $[t_{j-1, \text{new}}^i, t_{j+1}^i]$ for each node $u_j^i$ in cascade graph $\mathcal{G}_i$ (except the first and last nodes). 

\subsection{Experimental Results and Analysis}\label{subsec:results}

Comparison results are shown in Table~\ref{tab:results}, including three kinds of evaluation protocols: supervised, linear evaluation on frozen features, and semi-supervised with fine-tuning. 
The experimental results show that our proposed \M~framework outperforms all the baselines by a large margin. 
In the following, we give detailed discussions w.r.t.: (i) answering three research questions; (ii) two notable observations; and (iii) four ablation studies.

\vpara{Usage of unlabeled data improves prediction with model distillation involved.}
In Table~\ref{tab:results}, the results suggest that introducing a large amount of unlabeled data into \M's pre-training stage without distillation decreases the prediction performance, especially when labeled data are few and the model is linear evaluated (i.e., parameters in the pre-training phase are frozen). We speculate that the degradation of performance is owing to the interference of unlabeled data in the latent space. With model distillation, this deficiency can be largely abated and improves over the fine-tuned model. Another benefit is that the model becomes more generalizable for knowledge transfer as the unlabeled data is again well incorporated during this distillation process (see Section~\ref{subsec:transfer}).

\vpara{The cascade graph data augmentation is effective for contrastive learning.}
Three augmentation strategies all improve the prediction performance. The improvement becomes larger when fine-tuning on only a small fraction of labeled cascades. 
The performance of {\verb|AugSIM|} strategy is better than {\verb|AugRWR|} and \texttt{AugAttr}, which suggests that simulating the information diffusion in networks is a promising direction for graph contrastive learning on cascade modeling and prediction, compared to random walk-based or heuristic models. Specifically, when fine-tuned on 1\%, 10\%, and 100\% of the labeled data, \M~improves the Base model by approximately 5.3\%, 9.3\%, and 1.4\%, respectively, in terms of MSLE. This further supports that data augmentation capturing some variation and uncertainty of cascading is useful and can potential alleviate overfitting.

\begin{figure*}[t]
    \centering
    \includegraphics[width=0.95\linewidth]{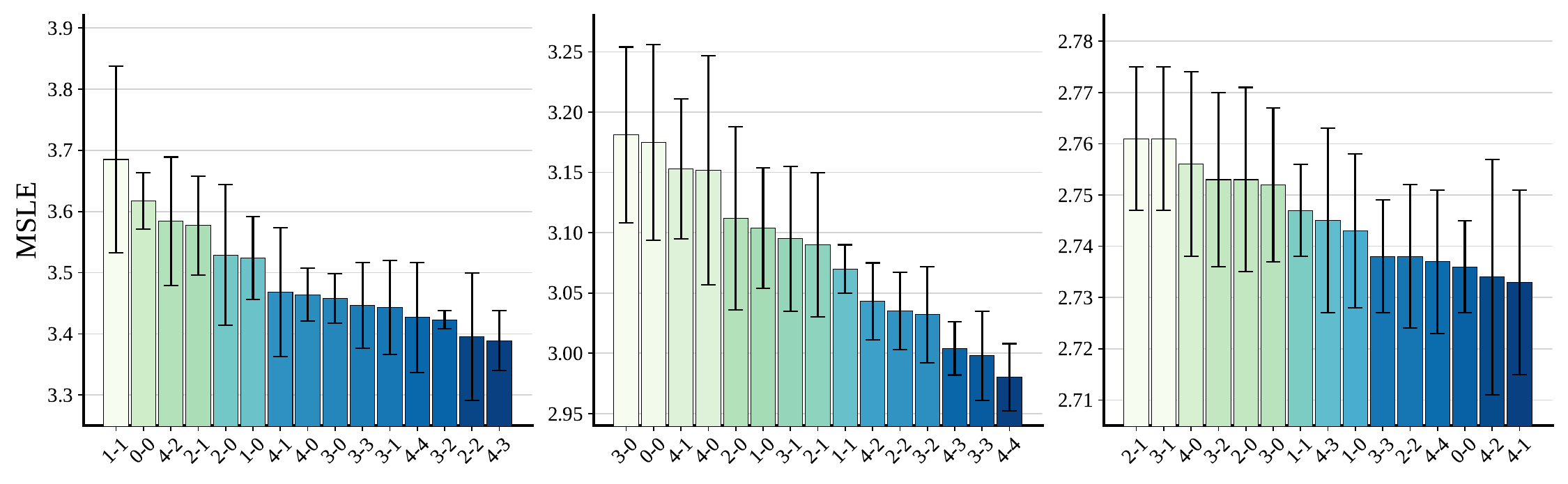}
    
    \caption{Impact of MLP-based projection head designs with different label fractions on the Weibo dataset. The labels of X-axis $i$-$j$ denotes that the projection head depth is $i$ and we fine-tune the \M~framework from the $j$-th layer of projection head. Mean MSLEs are reported by 5 runs with standard deviation. Left: 1\% labels; Middle: 10\% labels; Right: 100\% labels.
    }
    \label{fig:projection-head}
\end{figure*}

\begin{figure}[t]
    \centering
    \includegraphics[width=1.07\linewidth]{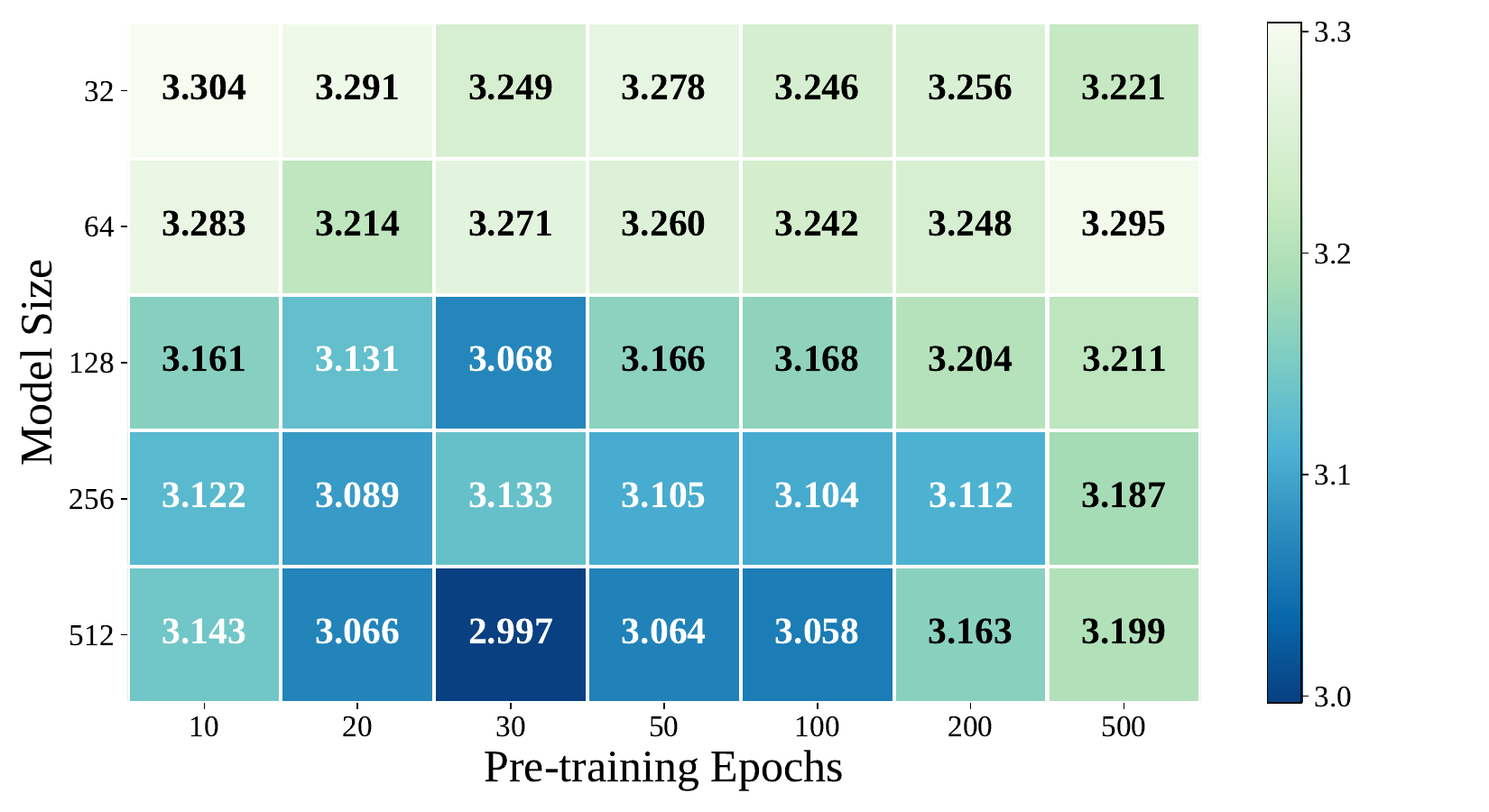}

    \caption{Impact of different combinations of model sizes and pre-training epochs of \M~(fine-tuned on 10\% of Weibo labeled cascades). Mean MSLEs are reported by 5 runs.
    }
    \label{fig:model-batch-size}
\end{figure}

\vpara{The \M~model performs on par with or even beats strong supervised counterparts.}
With all components we described in Section~\ref{sec:methodology} equipped and combined, i.e., with unlabeled cascade data, graph data augmentation strategy {\verb|AugSIM|} specifically designed for cascades, contrastive graph self-supervised pre-training, and task-specific fine-tuning and model distillation, our proposed \M~framework achieves a new state-of-the-art for cascade popularity prediction, outperforming a strong supervised \textbf{Base} model up to 9.2\%, 11.7\% and 2.9\% when fine-tuned on 1\%, 10\%, and 100\% of labeled data, respectively, in terms of MSLE on the Weibo dataset.

Furthermore, we have several notable observations.

\vpara{Observation 1: \M~is label-efficient compared to baselines.}
When fine-tuned on different fractions of labeled cascades, \M~is more label-efficient compared to the supervised model. With only 1\% of labels available, the performance of \M~is on par with Base trained on 10\% of labels ($3.25$ vs. $3.24$). This can be explained by the contrastive learning and data augmentation in~\M.

\vpara{Observation 2: Data augmentation does not benefit supervised learning for cascade prediction.} When original cascade graphs are augmented by {\verb|AugSIM|}, we observe that there is no benefit of graph data augmentation for supervised cascade learning. Actually, introducing {\verb|AugSIM|} into the Base model substantially lowers the prediction performance. The gap becomes larger when more labeled cascades are involved. One plausible reason is that supervised models learn feature-level representations rather abstract-level semantics, which are not able to capture variations and uncertainties brought in by augmentation. 

To demonstrate the robustness and sensitivity of \M, we perform several ablation studies.

\begin{table}[t]
    \centering
    \caption{
    Fine-tuning from a middle layer is not always the better choice.
    }
    \label{tab:which-layer}
    \begin{tabular}{lrrr}
    \toprule
        \textbf{Label Fraction} & \textbf{First Layer} & \textbf{Middle Layer(s)} & \textbf{Last Layer} \\ \midrule
        1\%     & 3.484 & \textbf{3.470} & 3.479 \\
        10\%    & 3.127 & 3.060 & \textbf{3.011} \\
        100\%   & 2.741 & 2.736 & \textbf{2.730} \\
    \bottomrule
    \end{tabular}
\end{table}

\vpara{Ablation 1: Cascade self-supervised learning benefits more from a deeper projection head, especially when labeled data are few.} 
Introducing a non-linear and learnable MLP-based projection head (cf.~\cite{chen2020simple}) showed that this simple mechanism can provide significant improvement for visual representation learning, as verified in~\cite{chen2020improved}. Subsequently, in \cite{chen2020big} it was shown that a deeper projection head and fine-tuning from a middle layer are a more powerful approach. However, it is not clear whether this mechanism has a significant advantage for cascade learning and prediction. We experimented with 15 different projection heads in \M~with varied label fractions, and the results are shown in \figurename~\ref{fig:projection-head}. 
We tag each projection head design as $i$-$j$, where $i$ is the depth of the projection head from 0 to 4, and $j$ denotes that the \M~framework is fine-tuned from the $j$-th layer of the head. 
For different fractions of labeled cascades used for fine-tuning, the behavior of projection heads varies. 
When fine-tuning on 1\% and 10\% of labeled cascades, the projection head significantly improves the prediction performance, on 13 out of 14 projection head designs other than $0$-$0$ (i.e., without projection head), leading up to $\sim$6.3\% improvement on 1\% labels and $\sim$6.2\% improvement on 10\% labels. 
When labeled data expand (e.g., with 100\% labels), only 2 out of 14 designs improve the prediction, which provides a negative case for using the projection head on cascade contrastive learning. 
As for projection head depth, we found that deeper heads are more effective than shallow heads. 
There is no evidence that fine-tuning from a middle layer is significantly better (contrary to previous work \cite{chen2020big}), as shown in Table~\ref{tab:which-layer}.
Compared to visual representation learning, where projection head brings common performance improvement, whether to use projection heads (and what kind) still lacks a universal guidance, at least for cascade graph learning. 

\begin{table}[t]
    \centering
    \setlength{\tabcolsep}{4.7pt}
    \caption{Impact of model size on (semi-)supervised models. Mean MSLEs are reported by 5 runs with std on 10\% of Weibo labeled cascades.
    }
    \label{tab:model-size-sup-semi-sup-10}
    \begin{tabular}{@{}lrrrrr@{}}
    \toprule
    \textbf{Model} & \textbf{1$\times$} & \textbf{2$\times$} & \textbf{4$\times$} & \textbf{8$\times$} & \textbf{16$\times$} \\ \midrule
    Superv. & 3.32$_{\pm 0.02}$ & 3.29$_{\pm 0.11}$ & 3.21$_{\pm 0.15}$ & 3.16$_{\pm 0.16}$ & 3.30$_{\pm 0.06}$ \\ \arrayrulecolor[gray]{0.65} \midrule
    Semi-su. & 3.22$_{\pm 0.06}^{{\color{orange}+0.10}}$ & 3.21$_{\pm 0.09}^{{\color{orange}+0.08}}$ & 3.07$_{\pm 0.04}^{{\color{orange}+0.14}}$ & 3.09$_{\pm 0.04}^{{\color{orange}+0.07}}$ & 2.99$_{\pm 0.02}^{{\color{orange}+0.31}}$ \\ 
    \arrayrulecolor[gray]{0} \bottomrule
    \end{tabular}
\end{table}

\vpara{Ablation 2: Model size and pre-training epochs.} \figurename~\ref{fig:model-batch-size} shows the relationship between different combinations of model sizes and pre-training epochs. Here we denote 1$\times$ as the model width (embedding dimension, units of RNNs and MLPs), set to 32. The setting of other hyper-parameters are: batch size is 64, augmentation strategy is {\verb|AugSIM|}, augmentation strength $\eta$ is 0.1, temperature $\tau$ is 0.05, and projection head is $2$-$0$. 
From the results, we can see that a large model is essential to guarantee a satisfactory performance. When the model size is already large, pre-training longer does not provide additional improvement, and even decreasing the performance, which might be due to that negative pre-training happens. The best performance is achieved by pre-training \M~(16$\times$) for 30 epochs.
We also investigated the impact of model size on supervised and semi-supervised models in Table~\ref{tab:model-size-sup-semi-sup-10}. 
From the results, we can see that for all model sizes, semi-supervised model outperforms supervised counterpart. As model size grows, supervised model is prone to overfitting, whereas semi-supervised model largely alleviates this due to two possible reasons: (i) incorporation of unlabeled data and data augmentation make the model more generalizable; and (ii) unlike supervised models where fine-grained feature-level representations are learned, the contrastive learning paradigm learns high-level abstract semantic representations. 

\begin{table}[t]
    \caption{MSLE of \M~trained on different distillation settings for teacher and student networks on Weibo dataset.
    }
    \label{tab:distill}
    
    \begin{center}
    \begin{tabular}{@{}llrrr@{}}
    \toprule
        \textbf{Teacher} & \textbf{Student} & \textbf{1\%} & \textbf{10\%} & \textbf{100\%} \\ \midrule
        {\color{gray}Base}            & {\color{gray}-}                             & {\color{gray}3.58} & {\color{gray}3.24} & {\color{gray}2.77} \\
        \rowcolor[gray]{.95}fine-tuned      & (w/o distillation)            & 3.39 & 2.94 & 2.73 \\
        label           & label (self-distill)          & 3.28 & 2.88 & 2.70 \\
        label+unlabel   & label (self-distill)          & 3.27 & 2.87 & 2.71 \\
        label+unlabel   & unlabel (self-distill)        & 3.29 & 2.89 & 2.84 \\
        label+unlabel   & label+unlabel (distilled)     & 3.26 & 2.88 & 2.77 \\
        \rowcolor[gray]{.95}label+unlabel   & label+unlabel (self-distill)  & \textbf{3.25} & \textbf{2.86} & \textbf{2.69} \\
    \bottomrule
    \end{tabular}
    \end{center}
\end{table}

\begin{figure}[t]
    \centering
    \includegraphics[width=\linewidth]{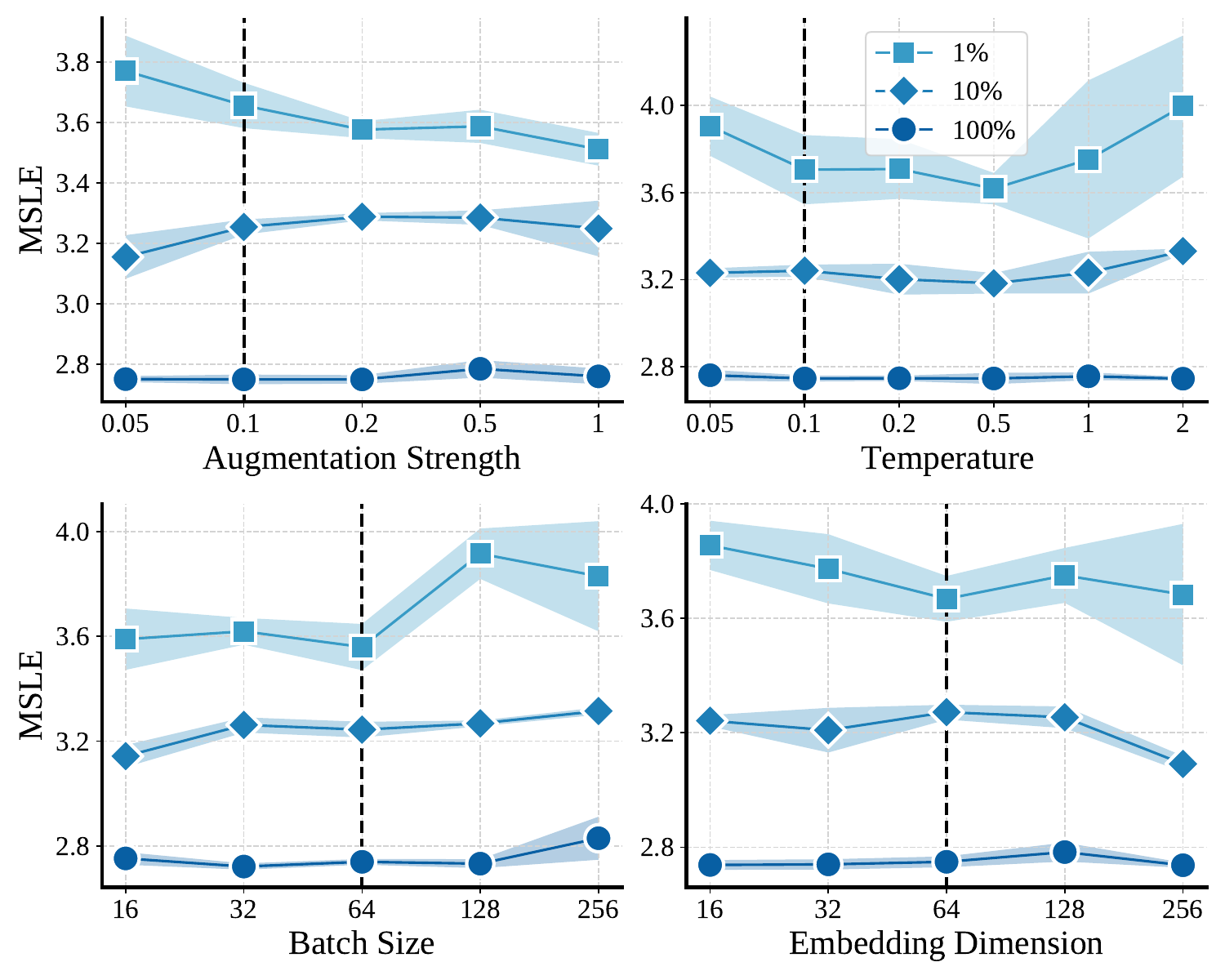}

    \caption{Sensitive analysis of hyper-parameters on Weibo with different label fractions. Vertical dashed lines indicate default settings used in experiments. Mean MSLEs are reported by 5 runs with standard deviation.}
    \label{fig:params}
\end{figure}

\vpara{Ablation 3: Model distillation with labeled and unlabeled cascades brings additional improvement over the fine-tuned model.}
We conduct experiments with two different teacher networks and four different student networks for model distillation. The results are shown in Table~\ref{tab:distill}. The teacher network is pre-trained with labeled cascades, or both labeled and unlabeled cascades. The student network is distilled with (i) labeled; (ii) unlabeled; or (iii) both labeled and unlabeled cascades. 
From the results we can see that model distillation indeed provides non-trivial additional performance improvement compared to contrastive learning, as much as 73.7\% relative improvement in MSLE by using 1\% of labels (26.7\% for 10\%, and 50.0\% for all labels). One plausible explanation might be that the distillation makes the model more task-agnostic and alleviate the issue of negative transfer.

\vpara{Ablation 4: Hyper-parameter sensitivity analysis.} 
Here we study several hyper-parameters and their impacts on cascade prediction performance through the Weibo dataset with varied label fractions. The results are shown in Fig.~\ref{fig:params}. We use the following parameter settings by default: batch size is 64, augmentation strategy is {\verb|AugSIM|}, augmentation strength $\eta$ is 0.1, temperature $\tau$ is 0.05, pre-training epochs is 100, embedding dimension is 64, model size is 64 (2$\times$), projection head is two fully-connected layers and we fine-tune the model before the projection head (i.e., $2$-$0$).

\noindent$\bullet$ \textit{Impact of augmentation strength $\eta$}: while augmentation strength $\eta$ controls the number of added/removed nodes in cascade graphs, we observe that strong augmentations are better when pre-trained on 1\% of the labels. 

\noindent$\bullet$ \textit{Impact of contrastive loss temperature $\tau$}: the results show that a small value of temperature (around 0.2) is preferred. Large temperatures (e.g., 1 or 2) sometimes make the model unable to converge (8 of 30 trials do not converge).

\noindent$\bullet$ \textit{Impact of batch size $B$}: contrary to previous conclusions \cite{chen2020simple} for contrastive learning on visual representation, learning of cascade representation prefers a smaller batch size. When full set of labeled cascades is used, the results with larger batch sizes (128 and 256) become worse and unstable. When 100$\times$ fewer labeled data are used, the model performs significantly worse. 
This suggests that a large batch size is not always necessary in contrastive learning. We conjecture that this deficiency is because the information and variances in cascades are significantly less than those in images (e.g., ImageNet), thus a smaller batch is sufficient for the model to distinguish different cascades. 
Such trends are also reported in \cite{qiu2020gcc}, where building a large batch (a.k.a. dictionary size) is not always helpful, and sometimes decreases the performance. However, a much larger batch (e.g., 4096/8192), or a momentum mechanism \cite{he2020momentum}, may improve the prediction when more (and complicated) cascade features are available, e.g., global user/item graph, texts and images of cascades. We leave this hypothesis for future work.

\noindent$\bullet$ \textit{Impact of embedding dimension}: we vary the embedding dimension from 16 to 256. The results 
indicate that a larger dimension is good for prediction. 
It also brings more time/space consumption and consequently makes the model prone to overfitting when labeled data are very few.

\begin{table}[t]
    \centering
    \renewcommand{\arraystretch}{1.25}
    \caption{Transferring knowledge across datasets. Abbrev.: Fine-tuned (FT). Mean MSLEs are reported by 10 runs with standard deviations. 
    }
    \label{tab:transfer-dataset}
    \begin{tabular}{@{}llrrr@{}}
    \toprule
        \textbf{Pre-trained on} & \textbf{FT on} & \textbf{1\%} & \textbf{10\%} & \textbf{100\%} \\ \midrule
        {\color{gray}Rand. init. (Weibo)} & {\color{gray}-} & {\color{gray}3.58$_{\pm 0.03}$} & {\color{gray}3.24$_{\pm 0.09}$} & {\color{gray}2.77$_{\pm 0.04}$} \\ 
        {\color{gray}Rand. init. (Twitter)} & {\color{gray}-} & {\color{gray}8.32$_{\pm 0.28}$} & {\color{gray}6.31$_{\pm 0.15}$} & {\color{gray}5.32$_{\pm 0.05}$} \\ 
        Weibo & Weibo & 3.39$_{\pm 0.10}^{{\color{orange}+0.19}}$ & 2.94$_{\pm 0.03}^{{\color{orange}+0.30}}$ & 2.73$_{\pm 0.02}$ \\
        \rowcolor[gray]{.95}Twitter & Twitter & 7.49$_{\pm 0.23}^{{\color{orange}+0.83}}$ & 5.81$_{\pm 0.06}^{{\color{orange}+0.50}}$ & 5.20$_{\pm 0.05}^{{\color{orange}+0.12}}$ \\
        Weibo & Twitter & 7.38$_{\pm 0.09}^{{\color{orange}+0.94}}$ & 5.77$_{\pm 0.06}^{{\color{orange}+0.54}}$ & 5.23$_{\pm 0.05}^{{\color{orange}+0.09}}$ \\ 
        Twitter & Weibo & 3.47$_{\pm 0.11}$ & 3.06$_{\pm 0.06}^{{\color{orange}+0.18}}$ & 2.75$_{\pm 0.02}$ \\ 
        Weibo \& Twitter & Weibo & 3.39$_{\pm 0.08}^{{\color{orange}+0.19}}$ & 2.98$_{\pm 0.02}^{{\color{orange}+0.26}}$ & \textbf{2.70}$_{\pm 0.02}$ \\ 
        \rowcolor[gray]{.95}& Twitter & \textbf{6.89$_{\pm 0.03}^{{\color{orange}+1.43}}$} & \textbf{5.70}$_{\pm 0.08}^{{\color{orange}+0.61}}$ & \textbf{5.14}$_{\pm 0.04}^{{\color{orange}+0.18}}$ \\ 
    \bottomrule
    \end{tabular}
\end{table}

\begin{table}[t]
    \centering
    \caption{Transferring knowledge to cascade outbreak prediction. Accuracy is reported on five balanced cascade datasets with varied label fractions. The \M~framework is pre-trained on the Weibo dataset for 30 epochs and fine-tuned on each of five other datasets. 
    }
    \label{tab:transfer-task}
    \vspace{-2mm}
    \begin{tabular}{@{}lllll@{}}
    \toprule
        \textbf{Data} & \textbf{Model} & \textbf{ACC (1\%)} & \textbf{ACC (10\%)} & \textbf{ACC (100\%)} \\ \midrule
        Weibo                   & rand. init.   & 82.64         & 82.37         & 83.70         \\
        \rowcolor[gray]{.95}    & \M            & 81.89 (-0.75) & 83.49 {\color{orange}(+1.12)} & 83.90 (+0.20) \\
        Twitter                 & rand. init.   & 61.50         & 78.44         & 84.06         \\
        \rowcolor[gray]{.95}    & \M            & 81.48 {\color{orange}(+19.9)} & 87.30 {\color{orange}(+8.86)}& 87.12 {\color{orange}(+3.06)} \\
        ACM                     & rand. init.   & 63.73         & 64.84         & 69.35         \\
        \rowcolor[gray]{.95}    & \M            & 60.49 {\color{orange}(-3.24)} & 68.86 {\color{orange}(+4.02)}& 71.09 {\color{orange}(+1.74)} \\
        APS                     & rand. init.   & 74.40         & 76.30         & 80.05         \\
        \rowcolor[gray]{.95}    & \M            & 77.28 {\color{orange}(+2.88)} & 79.94 {\color{orange}(+3.64)}& 81.10 {\color{orange}(+1.05)} \\
        DBLP                    & rand. init.   & 74.60         & 75.40         & 75.30         \\
        \rowcolor[gray]{.95}    & \M            & 74.16 (-0.44) & 76.96 {\color{orange}(+1.56)} & 77.28 {\color{orange}(+1.98)} \\
    \bottomrule
    \end{tabular}
\end{table}

\subsection{Knowledge Transfer}\label{subsec:transfer}

To investigate the generalization capability of our model, we explore the transferring ability of \M~on five information cascade datasets (Weibo, Twitter, ACM, APS, and DBLP) and two cascade prediction tasks (popularity prediction and outbreak prediction).

\vpara{Transferring knowledge across different cascade datasets.} 
To demonstrate that the representations learned by \M~have general transferable knowledge across cascade datasets, we conduct the following experiments on Weibo and Twitter datasets: (i) pre-train on one dataset and fine-tune on another dataset; (ii) pre-train on both datasets and fine-tune on one dataset. The results are shown in Table~\ref{tab:transfer-dataset}. 
We have two notable findings: (i) \M~pre-trained on Weibo and then fine-tuned on Twitter, significantly outperforms random initialized Base model by large margins. When labeled data are few, its performance also surpasses the model which is both pre-trained and fine-tuned on Twitter. This suggests that not only the Weibo dataset helps the Twitter cascade predictions, but also provides a better starting point for fine-tuning compared to Twitter dataset itself; and (ii) when pre-trained both on Weibo \& Twitter, fine-tuned \M~model achieves even better prediction performance compared to other combinations, up to 72.3\% relative improvement (1.43 vs. 0.83) on only 1\% of the Twitter labels. This might be because the model learns the generic knowledge in pre-training and task-specific knowledge in fine-tuning and distillation.

\begin{table*}[t]
    \setlength{\tabcolsep}{4.5pt}
    \centering
    \scriptsize
    \caption{Performance comparison of \M~and baselines on Twitter (T), ACM (A), APS (AP), DBLP (D) datasets, measured by 5 runs of mean MSLEs and std with varied label fractions (1\%, 10\%, 100\%). We use a self-distilled \M~trained on labeled and unlabeled cascades for 30 epochs.
    }
    \label{tab:results-on-other-datasets}
    \begin{tabular}{@{}lrrrrrrrrrrrr@{}}
    \toprule
        \textbf{Model} & \textbf{T (1\%)} & \textbf{T (10\%)} & \textbf{T (Full)} & \textbf{A (1\%)} & \textbf{A (10\%)} & \textbf{A (Full)} & \textbf{AP (1\%)} & \textbf{AP (10\%)} & \textbf{AP (Full)} & \textbf{D (1\%)} & \textbf{D (10\%)} & \textbf{D (Full)} \\ \midrule
        Feature & - & 7.97$_{\pm 0.32}$ & 6.78$_{\pm 0.52}$ & 1.41$_{\pm 0.02}$ & 1.30$_{\pm 0.01}$ & 1.25$_{\pm 0.01}$ & 2.42$_{\pm 0.11}$ & 1.88$_{\pm 0.01}$ & 1.82$_{\pm 0.01}$ & 2.31$_{\pm 0.12}$ & 2.12$_{\pm 0.02}$ & 2.01$_{\pm 0.03}$ \\
        \rowcolor[gray]{.95}Base & 8.32$_{\pm 0.28}$ & 6.31$_{\pm 0.15}$ & 5.32$_{\pm 0.05}$ & 1.47$_{\pm 0.00}$ & 1.27$_{\pm 0.00}$ & 1.23$_{\pm 0.01}$ & 2.53$_{\pm 0.03}$ & 1.85$_{\pm 0.03}$ & \textbf{1.78}$_{\pm 0.01}$ & 2.57$_{\pm 0.02}$ & 2.08$_{\pm 0.03}$ & \textbf{1.93}$_{\pm 0.02}$ \\ \midrule
        \M-FT & 7.49$_{\pm 0.23}$ & 5.81$_{\pm 0.06}$ & 5.20$_{\pm 0.05}$ & 1.42$_{\pm 0.01}$ & 1.26$_{\pm 0.01}$ & \textbf{1.21}$_{\pm 0.00}$ & 2.19$_{\pm 0.04}$ & 1.88$_{\pm 0.02}$ & 1.80$_{\pm 0.01}$ & 2.29$_{\pm 0.04}$ & 2.03$_{\pm 0.03}$ & 1.94$_{\pm 0.02}$ \\ 
        \rowcolor[gray]{.95}\M & \textbf{7.03}$_{\pm 0.06}$ & \textbf{5.75}$_{\pm 0.04}$ & \textbf{5.12}$_{\pm 0.02}$ & \textbf{1.42}$_{\pm 0.00}$ & \textbf{1.24}$_{\pm 0.00}$ & \textbf{1.22}$_{\pm 0.00}$ & \textbf{2.08}$_{\pm 0.02}$ & \textbf{1.81}$_{\pm 0.00}$ & \textbf{1.78}$_{\pm 0.00}$ & \textbf{2.18}$_{\pm 0.02}$ & \textbf{2.00}$_{\pm 0.02}$ & \textbf{1.92}$_{\pm 0.00}$ \\
        \rowcolor[gray]{.95}\footnotesize{(improves)} & 15.5\%$\uparrow$ & 8.9\%$\uparrow$ & 3.8\%$\uparrow$ & 3.4\%$\uparrow$ & 2.4\%$\uparrow$ & 0.1\%$\uparrow$ & 17.8\%$\uparrow$ & 2.1\%$\uparrow$ & - & 15.2\%$\uparrow$ & 3.8\%$\uparrow$ & 0.5\%$\uparrow$  \\
    \arrayrulecolor[gray]{0}
    \bottomrule
    \end{tabular}
\end{table*}

\vpara{Transferring knowledge to another prediction task.} 
In addition to the task of popularity prediction, we investigate the capability of knowledge transfer to different tasks using another downstream prediction task, cascade outbreak prediction.
For each of the five datasets, we select top 10\% cascades as outbreak and others as non-outbreak (i.e., negative instances). Since the distribution of cascade popularity is highly skewed, we undersampling the non-outbreak cascades to create a balanced dataset. The process of knowledge transfer is as follows. \M~is pre-trained on the Weibo dataset and all its hyper-parameters are fixed, which allows us to transfer knowledge without additional hyper-parameter tuning. However, this may lower the prediction accuracy. Overall, \M~achieves comparable performance to the randomly initialized Base model if not better across all five datasets. The results in Table~\ref{tab:transfer-task} give us the following two observations: (i) when only 1\% of labeled data is used, \M~substantially outperforms the supervised Base model by $19.9$ in accuracy on the Twitter dataset; (ii) although \M~is pre-trained only on the Weibo cascades, it performs well on other datasets, including both in social and scientific scenarios. This suggests that the knowledge pre-trained from Weibo cascades are successfully transferred across different datasets. In summary, \M~outperforms the Base model on 12 out of 15 experiments. We believe this transferring capability can be attributed to the general knowledge learned during the pre-training with unlabeled data in the contrastive learning framework, as well as the model distillation under the teacher-student framework.

\begin{table}[t]
    \setlength{\tabcolsep}{4.7pt}
    \footnotesize
    \begin{center}
    \renewcommand{\arraystretch}{1.17}
    \caption{Impact of model size on (semi-)supervised models. Mean MSLEs are reported by 10 runs with std on 1\% or 100\% of Weibo labeled data.
    }
    \label{tab:model-size-sup-semi-sup}
    \begin{tabular}{@{}lrrrrr@{}}
    \toprule
    \textbf{Model} & \textbf{1$\times$} & \textbf{2$\times$} & \textbf{4$\times$} & \textbf{8$\times$} & \textbf{16$\times$} \\ \midrule
    \multicolumn{6}{@{}l}{\textit{trained with 1\% labeled cascades:}} \\
    Superv. & 3.74$_{\pm 0.24}$ & 3.57$_{\pm 0.08}$ & 3.98$_{\pm 0.35}$ & 4.44$_{\pm 0.07}$ & 4.53$_{\pm 0.04}$ \\ 
    Semi-su. & 3.73$_{\pm 0.04}$ & 3.52$_{\pm 0.07}^{{\color{orange}+0.05}}$ & 3.47$_{\pm 0.03}^{{\color{orange}+0.52}}$ & 3.45$_{\pm 0.10}^{{\color{orange}+1.01}}$ & 3.51$_{\pm 0.05}^{{\color{orange}+1.02}}$ \\ \midrule
    \multicolumn{6}{@{}l}{\textit{trained with 100\% labeled cascades:}} \\
    Superv. & 2.78$_{\pm 0.05}$ & 2.79$_{\pm 0.04}$ & 2.83$_{\pm 0.03}$ & 2.78$_{\pm 0.02}$ & 2.83$_{\pm 0.03}$ \\ 
    Semi-su. & 2.77$_{\pm 0.05}$ & 2.73$_{\pm 0.02}^{{\color{orange}+0.06}}$ & 2.75$_{\pm 0.01}^{{\color{orange}+0.08}}$ & 2.75$_{\pm 0.01}$ & 2.81$_{\pm 0.05}$ \\ 
    \bottomrule
    \end{tabular}
    \end{center}
\end{table}

\begin{figure}[t]
    \centering
    \subfloat[Representation $\mathbf{h}$]{\includegraphics[width=0.243\textwidth]{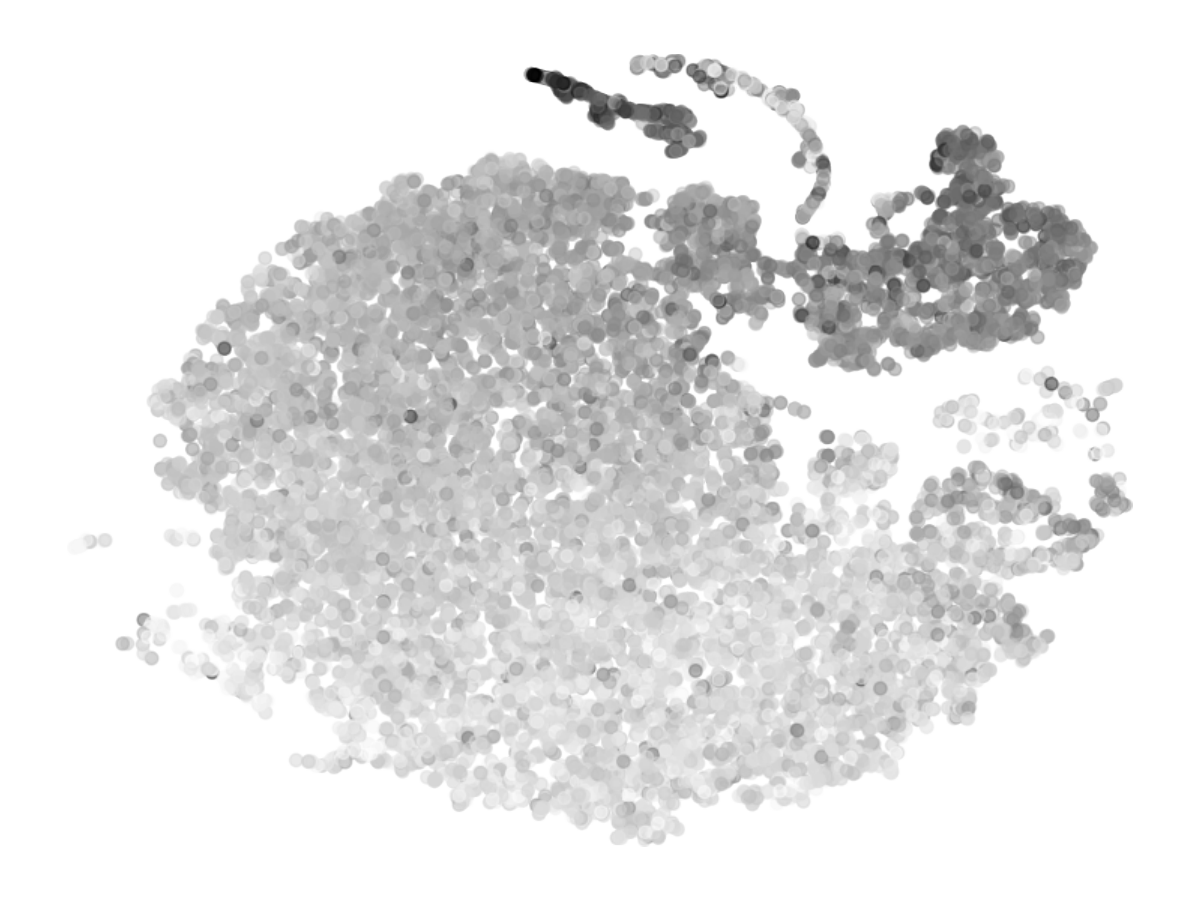}}
    \subfloat[Representation $\mathbf{z}$]{\includegraphics[width=0.243\textwidth]{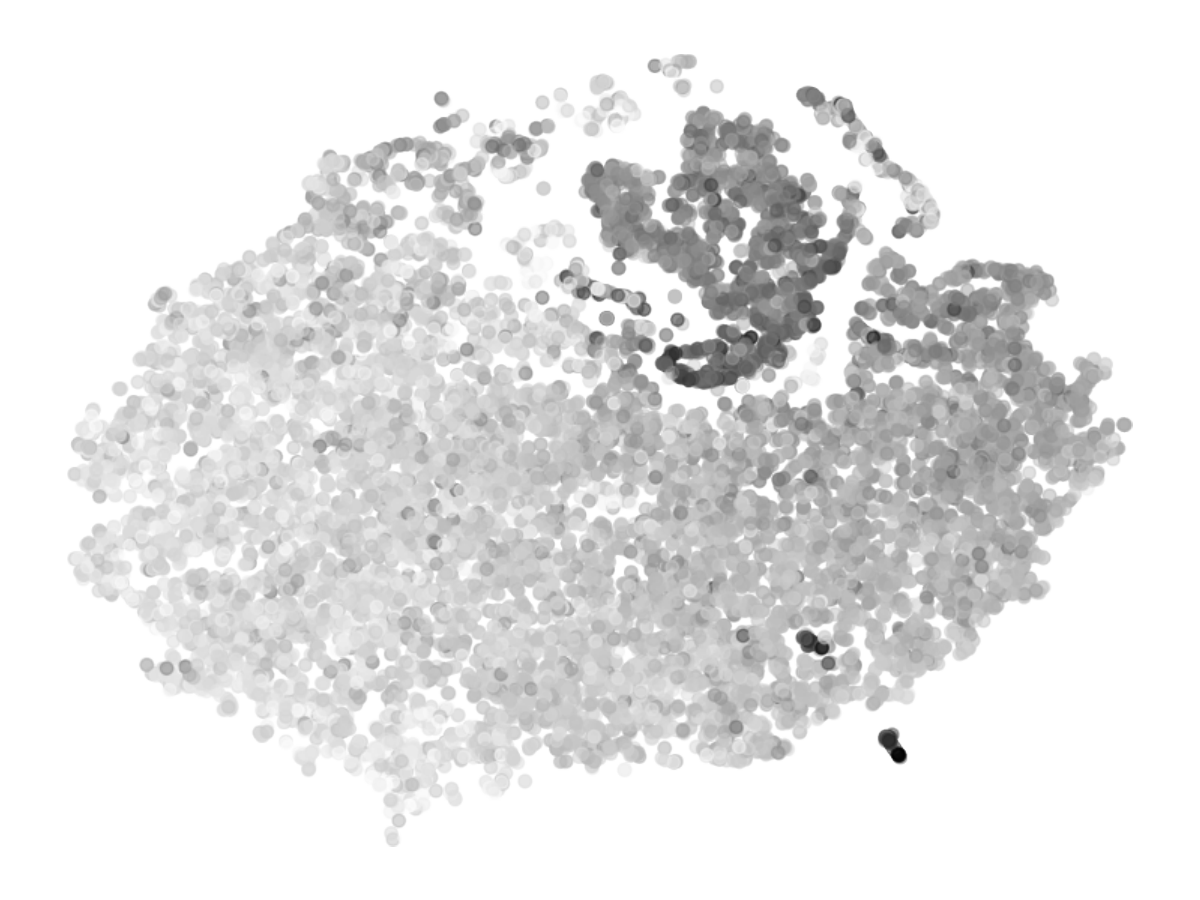}}
    
    \subfloat[$\mathbf{h}$ from supervised Base]{\includegraphics[width=0.243\textwidth]{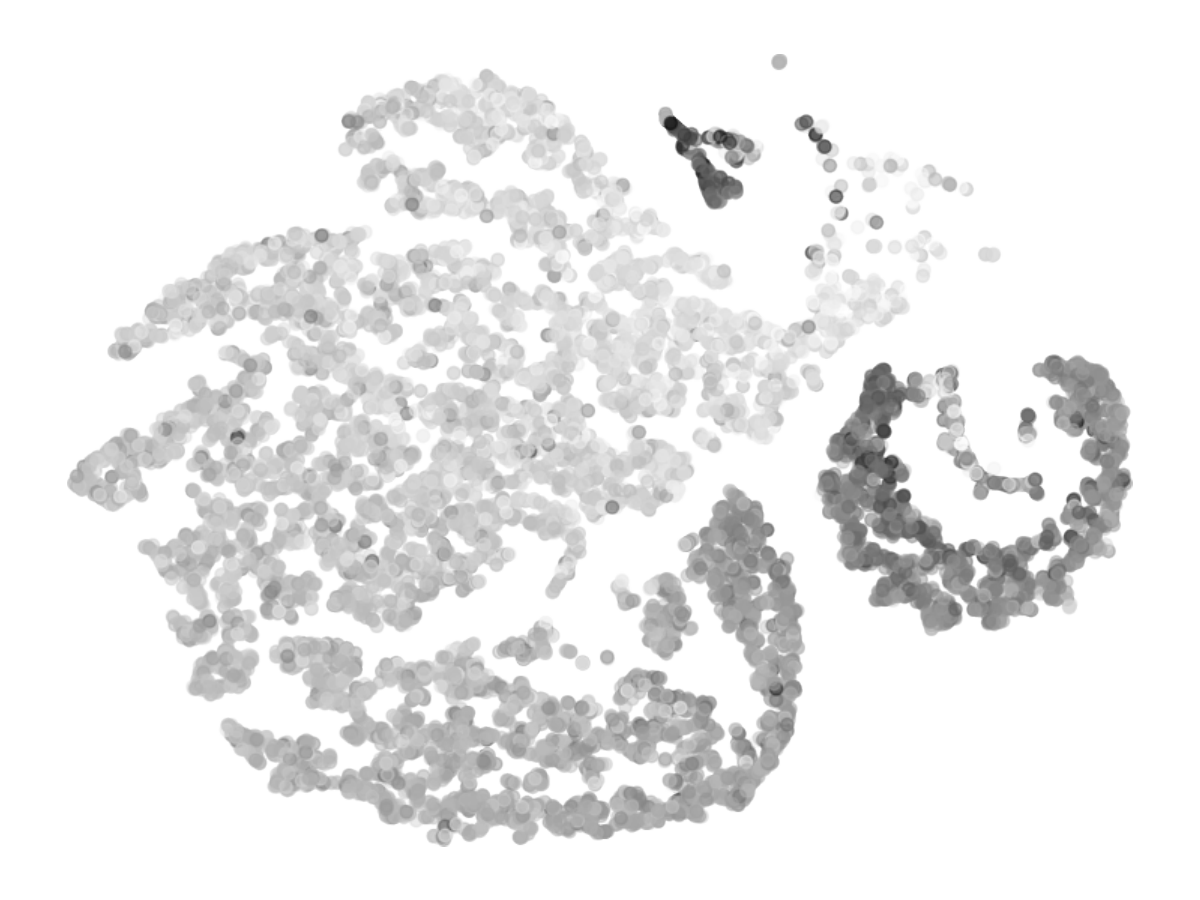}}
    \subfloat[$\mathbf{h}$ from linear evaluated \M]{\includegraphics[width=0.243\textwidth]{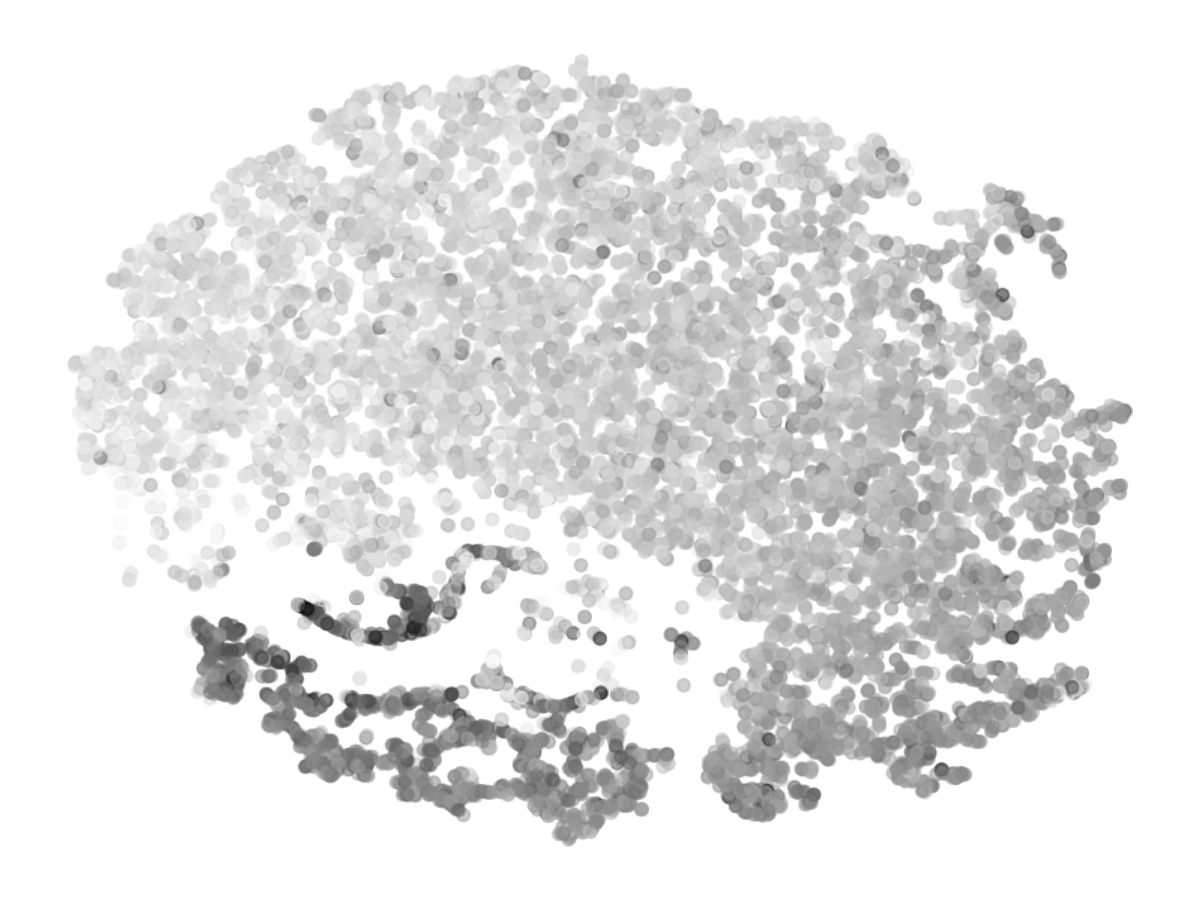}}
    
    \subfloat[$\mathbf{h}$ from fine-tuned \M]{\includegraphics[width=0.243\textwidth]{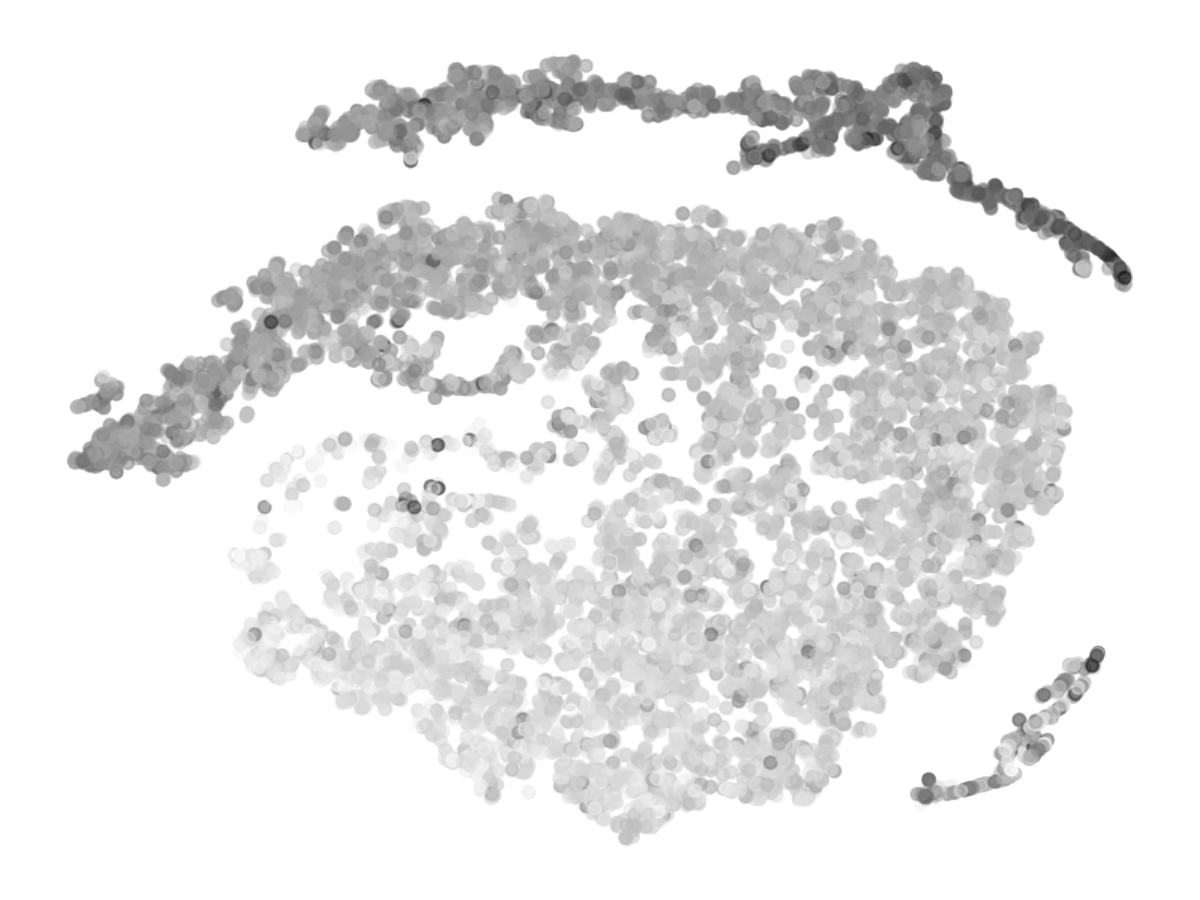}}
    \subfloat[$\mathbf{h}$ from distilled \M]{\includegraphics[width=0.243\textwidth]{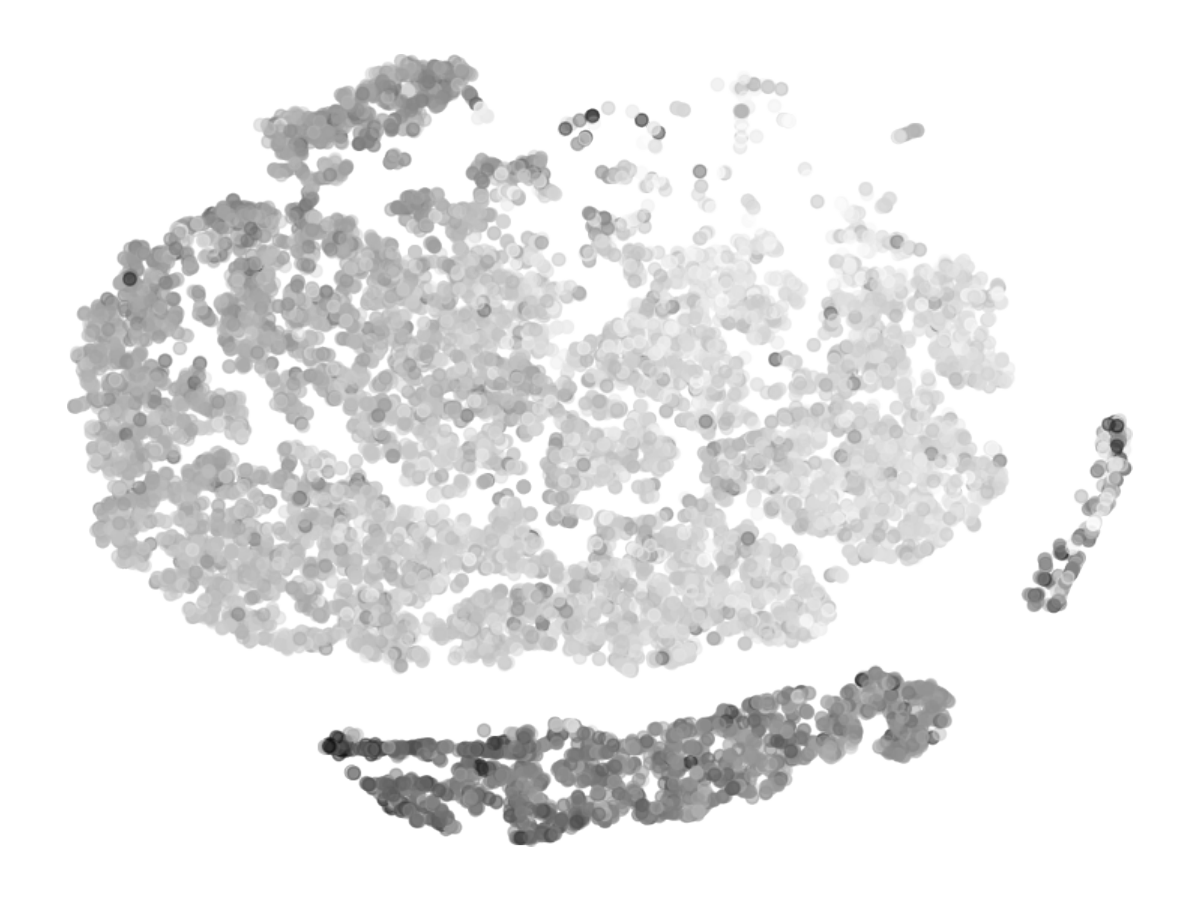}}

    \caption{Visualization of latent representations using $t$-SNE. (a) and (b) are $\mathbf{h}$ and $\mathbf{z}$ from pre-trained \M~model on 19,538 Weibo training cascades. (c) to (f) are latent vector $\mathbf{h}$ from Base, linear evaluated, fine-tuned, or distilled \M, respectively, on 15,630 Weibo test cascades.
    }
    \label{fig:tsne}
\end{figure}

\subsection{Additional Observations}\label{subsec:observations}

\vpara{Visualization of latent representations.}
In Fig.~\ref{fig:tsne} we visualize the learned representations on Weibo cascades using $t$-SNE. The first two plots (a) and (b) are representations $\mathbf{h}$ and $\mathbf{z}$ pre-trained by \M~on 19,538 Weibo cascades (without label information). Recall that we use $\mathbf{h}$ for downstream prediction tasks and $\mathbf{z}$ for contrastive loss. The projection head is $4$-$1$, the embedding dimension of latent vectors is 256. The last four plots are representations $\mathbf{h}$ retrieved from (c) supervised Base model; (d) linear evaluated \M; (e) fine-tuned \M; and (f) distilled \M. Models are trained on 10\% of the labels. The color of each point (i.e., cascade) indicates its future popularity, where darker ones are larger cascades. We can see that compared to pre-trained or linear-evaluated $\mathbf{h}$, the visualization of \textit{task-specific} fine-tuned/distilled $\mathbf{h}$ is more separable between small and large cascades, which indicates that the model fine-tuning and distillation is effective for cascade popularity prediction. The representations in (a), (b), and (d) are smooth and not distinguishable, which suggests that the pre-trained model carries over and the ``negative'' transfer might occur. The representation from the supervised Base model in (c) does not well cluster larger cascades together. This verifies its inferior performance due to the fact that it lacks of contrastive pre-training and model distillation.

\vpara{Impact of model size.}
In Table~\ref{tab:model-size-sup-semi-sup}, we report an additional ablation study on model size for 1\% or 100\% of the label fractions. We can see that for supervised models, when model becomes larger, the prediction performance decreases severely. On the other side, semi-supervised model which involves pre-training and fine-tuning, is stable and even better when model size is large. 

\vpara{Results on other datasets.}
In Table~\ref{tab:results-on-other-datasets} we show the prediction performance of distilled \M~on Twitter, ACM, APS, and DBLP datasets. 
Our proposed \M~model performs on par or better than the supervised Base model as well as feature-based model) on four datasets. The achieved improvements are generally larger when label data are fewer.

\section{Conclusion}
\label{sec:conclusion}

We presented \M, a contrastive cascade graph learning framework which provides a new perspective of modeling cascade graphs, bridging the gap between supervised and unsupervised information cascade modeling and prediction, and enabling data augmentation strategies. 
Our experiments conducted on five different information diffusion datasets and two cascade prediction tasks, demonstrate the effectiveness of the devised cascade graph data augmentation strategy and \M's \textit{contrastive self-supervised pre-training}, \textit{fine-tuning}, and \textit{distillation} paradigm. In addition to performance improvement, our method is capable of exploiting unlabeled data and extracting supervision signals in a self-supervised manner. 
We also showed that the proposed model is label-efficient and can generalize information across different cascade data and applications.

As our future work, there are several aspects of the proposed model that warrant further investigation: (i) learning cascade graph representations dynamically via temporal embedding techniques \cite{xu2020inductive,singer2019node}; (ii) combining multiple datasets into unsupervised pre-training for better knowledge transfer; (iii) other possible cascade graph augmentation strategies, such as introducing more cascade features into selection of nodes/edges, as well as alternative contrastive mechanisms, e.g., momentum updates \cite{he2020momentum}, autoregressive modeling \cite{oord2018representation}, mutual information maximization \cite{velickovic2019deep} or multi-view contrasting \cite{hassani2020contrastive}; and (iv) learning cascade representation in multi-modalities setting \cite{zhou2021survey}.

% or


% \appendices

% \section{}

%\ifCLASSOPTIONcompsoc
 %\section*{Acknowledgments}
%\else
%\section*{Acknowledgment}
%\fi
%{\footnotesize Work was supported by National Natural Science Foundation of China Grants No.~62072077, 61602097, and NSF Grants 1646107, 2030249.}

\ifCLASSOPTIONcaptionsoff
  \newpage
\fi

% \IEEEtriggeratref{8}


\bibliographystyle{IEEEtran}

\bibliography{IEEEabrv,./xovee}

\begin{thebibliography}{10}
\providecommand{\url}[1]{#1}
\csname url@samestyle\endcsname
\providecommand{\newblock}{\relax}
\providecommand{\bibinfo}[2]{#2}
\providecommand{\BIBentrySTDinterwordspacing}{\spaceskip=0pt\relax}
\providecommand{\BIBentryALTinterwordstretchfactor}{4}
\providecommand{\BIBentryALTinterwordspacing}{\spaceskip=\fontdimen2\font plus
\BIBentryALTinterwordstretchfactor\fontdimen3\font minus
  \fontdimen4\font\relax}
\providecommand{\BIBforeignlanguage}[2]{{%
\expandafter\ifx\csname l@#1\endcsname\relax
\typeout{** WARNING: IEEEtran.bst: No hyphenation pattern has been}%
\typeout{** loaded for the language `#1'. Using the pattern for}%
\typeout{** the default language instead.}%
\else
\language=\csname l@#1\endcsname
\fi
#2}}
\providecommand{\BIBdecl}{\relax}
\BIBdecl

\bibitem{zhou2021survey}
F.~Zhou, X.~Xu, G.~Trajcevski, and K.~Zhang, ``A survey of information cascade
  analysis: Models, predictions and recent advances,'' \emph{ACM Computing
  Surveys}, vol.~54, no.~2, 2021.

\bibitem{chang2021mobility}
S.~Chang, E.~Pierson, P.~W. Koh, J.~Gerardin, B.~Redbird, D.~Grusky, and
  J.~Leskovec, ``Mobility network models of covid-19 explain inequities and
  inform reopening,'' \emph{Nature}, vol. 589, no. 7840, pp. 82--87, 2021.

\bibitem{yang2021full}
C.~Yang, H.~Wang, J.~Tang, C.~Shi, M.~Sun, G.~Cui, and Z.~Liu, ``Full-scale
  information diffusion prediction with reinforced recurrent networks,''
  \emph{TNNLS}, 2021, 13 pages, in press.

\bibitem{li2017deepcas}
C.~Li, J.~Ma, X.~Guo, and Q.~Mei, ``Deep{C}as: An end-to-end predictor of
  information cascades,'' in \emph{WWW}, 2017, pp. 577--586.

\bibitem{tang2021fully}
X.~Tang, D.~Liao, W.~Huang, J.~Xu, L.~Zhu, and M.~Shen, ``Fully exploiting
  cascade graphs for real-time forwarding prediction,'' in \emph{AAAI}, 2021,
  pp. 582--590.

\bibitem{cao2017deephawkes}
Q.~Cao, H.~Shen, K.~Cen, W.~Ouyang, and X.~Cheng, ``Deep{H}awkes: Bridging the
  gap between prediction and understanding of information cascades,'' in
  \emph{CIKM}, 2017, pp. 1149--1158.

\bibitem{wu2020unsupervised}
M.~Wu, S.~Pan, C.~Zhou, X.~Chang, and X.~Zhu, ``Unsupervised domain adaptive
  graph convolutional networks,'' in \emph{WWW}, 2020, pp. 1457--1467.

\bibitem{kingma2013auto}
D.~P. Kingma and M.~Welling, ``Auto-encoding variational bayes,''
  \emph{arXiv:1312.6114}, 2013, 14 pages.

\bibitem{hu2020strategies}
W.~Hu, B.~Liu, J.~Gomes, M.~Zitnik, P.~Liang, V.~Pande, and J.~Leskovec,
  ``Strategies for pre-training graph neural networks,'' in \emph{ICLR}, 2020,
  22 pages.

\bibitem{cheng2014can}
J.~Cheng, L.~Adamic, P.~A. Dow, J.~M. Kleinberg, and J.~Leskovec, ``Can
  cascades be predicted?'' in \emph{WWW}, 2014, pp. 925--936.

\bibitem{zhao2018comparative}
Y.~Zhao, C.~Wang, C.-H. Chi, K.-Y. Lam, and S.~Wang, ``A comparative study of
  transactional and semantic approaches for predicting cascades on {T}witter,''
  in \emph{IJCAI}, 2018, pp. 1212--1218.

\bibitem{zhao2015seismic}
Q.~Zhao, M.~A. Erdogdu, H.~Y. He, A.~Rajaraman, and J.~Leskovec, ``{SEISMIC}: A
  self-exciting point process model for predicting tweet popularity,'' in
  \emph{KDD}, 2015, pp. 1513--1522.

\bibitem{gao2020using}
X.~Gao, X.~Jia, C.~Yang, and G.~Chen, ``Using survival theory in early pattern
  detection for viral cascades,'' \emph{TKDE}, 2020, 15 pages, in press.

\bibitem{zhou2020variational}
F.~Zhou, X.~Xu, K.~Zhang, G.~Trajcevski, and T.~Zhong, ``Variational
  information diffusion for probabilistic cascades prediction,'' in
  \emph{INFOCOM}, 2020, pp. 1618--1627.

\bibitem{cao2020popularity}
Q.~Cao, H.~Shen, J.~Gao, B.~Wei, and X.~Cheng, ``Popularity prediction on
  social platforms with coupled graph neural networks,'' in \emph{WSDM}, 2020,
  pp. 70--78.

\bibitem{he2020momentum}
K.~He, H.~Fan, Y.~Wu, S.~Xie, and R.~Girshick, ``Momentum contrast for
  unsupervised visual representation learning,'' in \emph{CVPR}, 2020, pp.
  9729--9738.

\bibitem{wu2018unsupervised}
Z.~Wu, Y.~Xiong, S.~X. Yu, and D.~Lin, ``Unsupervised feature learning via
  non-parametric instance discrimination,'' in \emph{CVPR}, 2018, pp.
  3733--3742.

\bibitem{misra2020self}
I.~Misra and L.~v.~d. Maaten, ``Self-supervised learning of pretext-invariant
  representations,'' in \emph{CVPR}, 2020, pp. 6707--6717.

\bibitem{bachman2019learning}
P.~Bachman, R.~D. Hjelm, and W.~Buchwalter, ``Learning representations by
  maximizing mutual information across views,'' in \emph{NeurIPS}, 2019, pp.
  15\,535--15\,545.

\bibitem{velickovic2019deep}
P.~Velickovic, W.~Fedus, W.~L. Hamilton, P.~Li{\`o}, Y.~Bengio, and R.~D.
  Hjelm, ``Deep graph infomax,'' in \emph{ICLR}, 2019, 17 pages.

\bibitem{oord2018representation}
A.~v.~d. Oord, Y.~Li, and O.~Vinyals, ``Representation learning with
  contrastive predictive coding,'' \emph{arXiv:1807.03748}, 2018, 13 pages.

\bibitem{chen2020big}
T.~Chen, S.~Kornblith, K.~Swersky, M.~Norouzi, and G.~Hinton, ``Big
  self-supervised models are strong semi-supervised learners,'' in
  \emph{NeurIPS}, 2020, 13 pages.

\bibitem{chen2020simple}
T.~Chen, S.~Kornblith, M.~Norouzi, and G.~Hinton, ``A simple framework for
  contrastive learning of visual representations,'' in \emph{ICML}, 2020, pp.
  1597--1607.

\bibitem{cubuk2019autoaugment}
E.~D. Cubuk, B.~Zoph, D.~Mane, V.~Vasudevan, and Q.~V. Le, ``Autoaugment:
  Learning augmentation strategies from data,'' in \emph{CVPR}, 2019, pp.
  113--123.

\bibitem{zhao2020data}
T.~Zhao, Y.~Liu, L.~Neves, O.~Woodford, M.~Jiang, and N.~Shah, ``Data
  augmentation for graph neural networks,'' in \emph{AAAI}, 2021, pp.
  11\,015--11\,023.

\bibitem{rong2019dropedge}
Y.~Rong, W.~Huang, T.~Xu, and J.~Huang, ``Drop{E}dge: Towards deep graph
  convolutional networks on node classification,'' in \emph{ICLR}, 2019, 17
  pages.

\bibitem{chen2020measuring}
D.~Chen, Y.~Lin, W.~Li, P.~Li, J.~Zhou, and X.~Sun, ``Measuring and relieving
  the over-smoothing problem for graph neural networks from the topological
  view,'' in \emph{AAAI}, 2020, pp. 3438--3445.

\bibitem{qiu2020gcc}
J.~Qiu, Q.~Chen, Y.~Dong, J.~Zhang, H.~Yang, M.~Ding, K.~Wang, and J.~Tang,
  ``{GCC}: Graph contrastive coding for graph neural network pre-training,'' in
  \emph{KDD}, 2020, pp. 1150--1160.

\bibitem{hassani2020contrastive}
K.~Hassani and A.~H. Khasahmadi, ``Contrastive multi-view representation
  learning on graphs,'' in \emph{ICML}, 2020, 13 pages.

\bibitem{ying2018transfer}
W.~Ying, Y.~Zhang, J.~Huang, and Q.~Yang, ``Transfer learning via learning to
  transfer,'' in \emph{ICML}, 2018, pp. 5085--5094.

\bibitem{hu2020gpt}
Z.~Hu, Y.~Dong, K.~Wang, K.-W. Chang, and Y.~Sun, ``{GPT-GNN}: Generative
  pre-training of graph neural networks,'' in \emph{KDD}, 2020, pp. 1857--1867.

\bibitem{guille2013information}
A.~Guille, H.~Hacid, C.~Favre, and D.~A. Zighed, ``Information diffusion in
  online social networks: A survey,'' \emph{ACM Sigmod Record}, vol.~42, no.~2,
  pp. 17--28, 2013.

\bibitem{chen2019information}
X.~Chen, F.~Zhou, K.~Zhang, G.~Trajcevski, T.~Zhong, and F.~Zhang,
  ``Information diffusion prediction via recurrent cascades convolution,'' in
  \emph{ICDE}, 2019, pp. 770--781.

\bibitem{sun2020infograph}
F.-Y. Sun, J.~Hoffman, V.~Verma, and J.~Tang, ``Info{G}raph: Unsupervised and
  semi-supervised graph-level representation learning via mutual information
  maximization,'' in \emph{ICLR}, 2020, 16 pages.

\bibitem{crane2008robust}
R.~Crane and D.~Sornette, ``Robust dynamic classes revealed by measuring the
  response function of a social system,'' \emph{PNAS}, vol. 105, no.~41, pp.
  15\,649--15\,653, 2008.

\bibitem{chen2020improved}
X.~Chen, H.~Fan, R.~Girshick, and K.~He, ``Improved baselines with momentum
  contrastive learning,'' \emph{arXiv:2003.04297}, 2020, 3 pages.

\bibitem{hjelm2018learning}
R.~D. Hjelm, A.~Fedorov, S.~Lavoie-Marchildon, K.~Grewal, P.~Bachman,
  A.~Trischler, and Y.~Bengio, ``Learning deep representations by mutual
  information estimation and maximization,'' in \emph{ICLR}, 2019, 24 pages.

\bibitem{weng2013virality}
L.~Weng, F.~Menczer, and Y.-Y. Ahn, ``Virality prediction and community
  structure in social networks,'' \emph{Scientific Reports}, vol.~3, 2013,
  article 2522, 3 pages.

\bibitem{tang2008arnetminer}
J.~Tang, J.~Zhang, L.~Yao, J.~Li, L.~Zhang, and Z.~Su, ``Arnet{M}iner:
  Extraction and mining of academic social networks,'' in \emph{KDD}, 2008, pp.
  990--998.

\bibitem{grover2016node2vec}
A.~Grover and J.~Leskovec, ``node2vec: Scalable feature learning for
  networks,'' in \emph{KDD}, 2016, pp. 855--864.

\bibitem{wang2020nodeaug}
Y.~Wang, W.~Wang, Y.~Liang, Y.~Cai, J.~Liu, and B.~Hooi, ``{NodeAug}:
  Semi-supervised node classification with data augmentation,'' in \emph{KDD},
  2020, pp. 207--217.

\bibitem{xu2020inductive}
D.~Xu, C.~Ruan, E.~Korpeoglu, S.~Kumar, and K.~Achan, ``Inductive
  representation learning on temporal graphs,'' in \emph{ICLR}, 2020, 19 pages.

\bibitem{singer2019node}
U.~Singer, I.~Guy, and K.~Radinsky, ``Node embedding over temporal graphs,'' in
  \emph{IJCAI}, 2019, pp. 4605--4612.

\end{thebibliography}


% H.~Kopka and P.~W. Daly, \emph{A Guide to {\LaTeX}}, 3rd~ed.\hskip 1em plus

% \end{thebibliography}


% \end{IEEEbiography}

% 1.25 to 1.0

\begin{IEEEbiography}[{\includegraphics[width=1in,height=1.25in,clip,keepaspectratio]{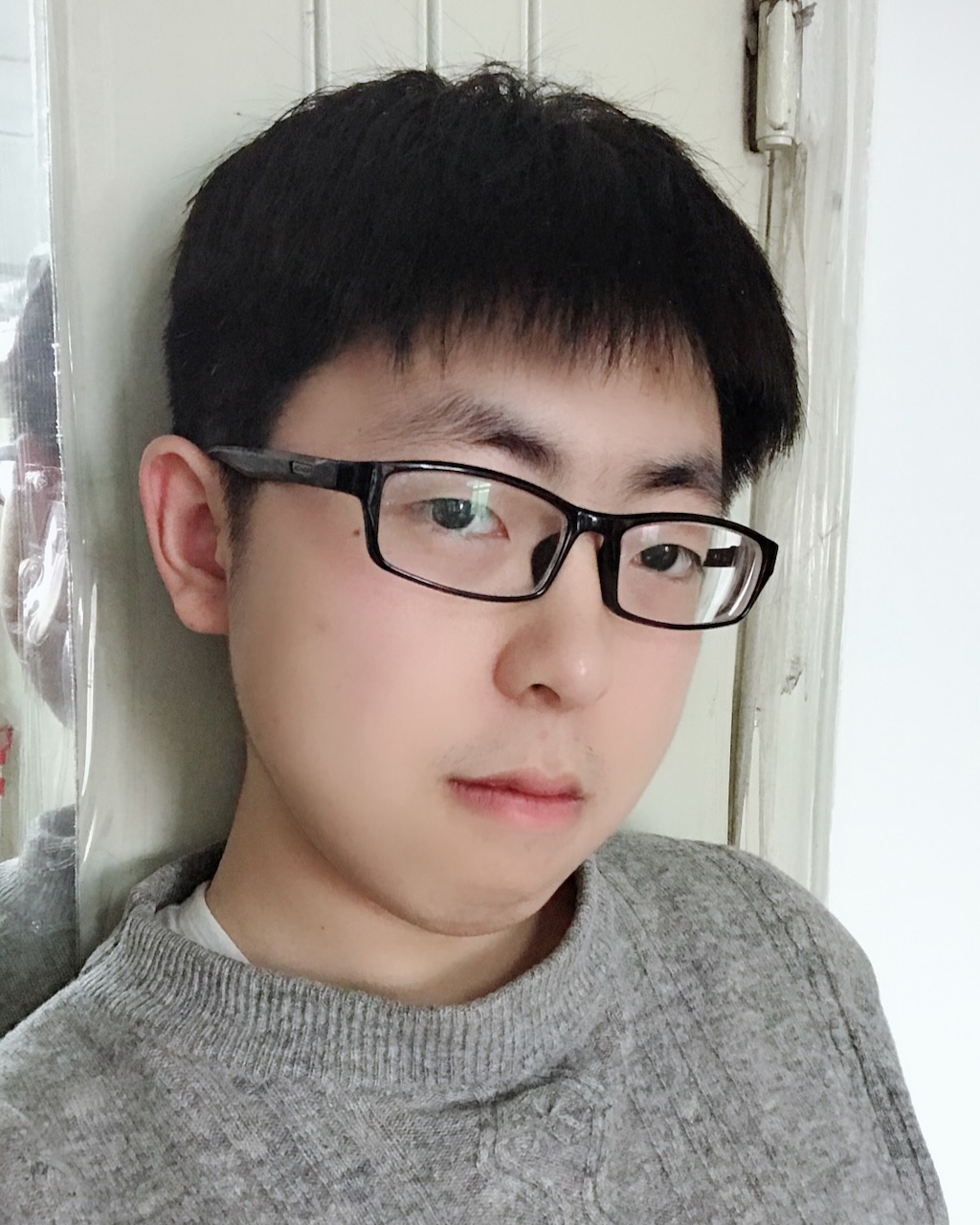}}]{Xovee Xu}
(GS’20) was born in Yulin, Shaanxi, China, in 1996. He received the B.S. degree and M.S. degree in software engineering from the University of Electronic Science and Technology of China (UESTC), Chengdu, Sichuan, China, in 2018 and 2021, respectively. He is currently pursuing the Ph.D. degree in computer science at UESTC. 

He is the author of several research articles published in INFOCOM, SIGIR, TKDE, AAAI, and CSUR. His recent research interests include social network data mining and knowledge discovery, primarily focuses on information diffusion in full-scale graphs, human-centered data mining, representation learning, and their novel applications in various social and scientific scenarios such as information cascade popularity prediction, urban flow inference, and scientific impact prediction.
\end{IEEEbiography}

\begin{IEEEbiography}[{\includegraphics[width=1in,height=1.25in,clip,keepaspectratio]{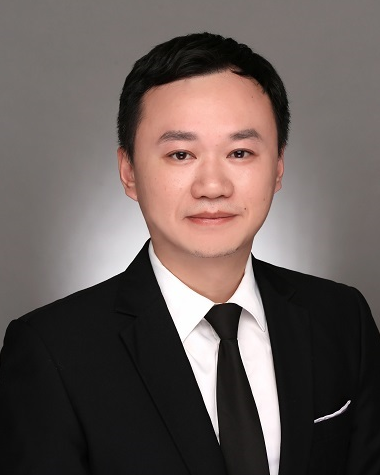}}]{Fan Zhou}
received the B.S. degree in computer science from Sichuan University, China, in 2003, and the M.S. and Ph.D. degrees from the University of Electronic Science and Technology of China, in 2006 and 2012, respectively, where he is currently an Associate Professor with School of Information and Software Engineering.

His research interests include machine learning, neural networks, spatio-temporal data management, graph learning, recommender systems, and social network mining and knowledge discovery.
\end{IEEEbiography}

\begin{IEEEbiography}[{\includegraphics[width=1in,height=1.25in,clip,keepaspectratio]{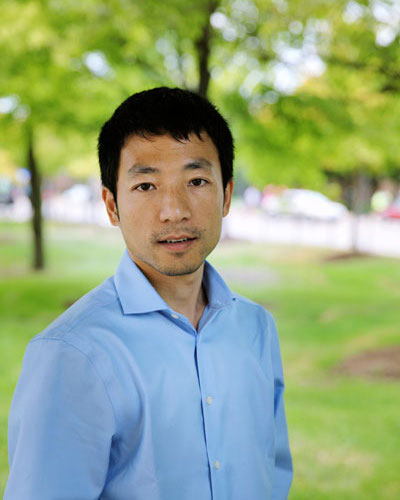}}]{Kunpeng Zhang} is Assistant Professor at Robert H. Smith School of Business, University of Maryland, College Park. He received the Ph.D. degree in computer science from Northwestern University, USA. He is interested in large-scale data analysis, with particular focuses on social data mining, image understanding via machine learning, social network analysis, and causal inference.

He has published papers in the area of social media, artificial intelligence, network analysis, and information systems on various conferences and journals.

\end{IEEEbiography}

\begin{IEEEbiography}[{\includegraphics[width=1in,height=1.25in,clip,keepaspectratio]{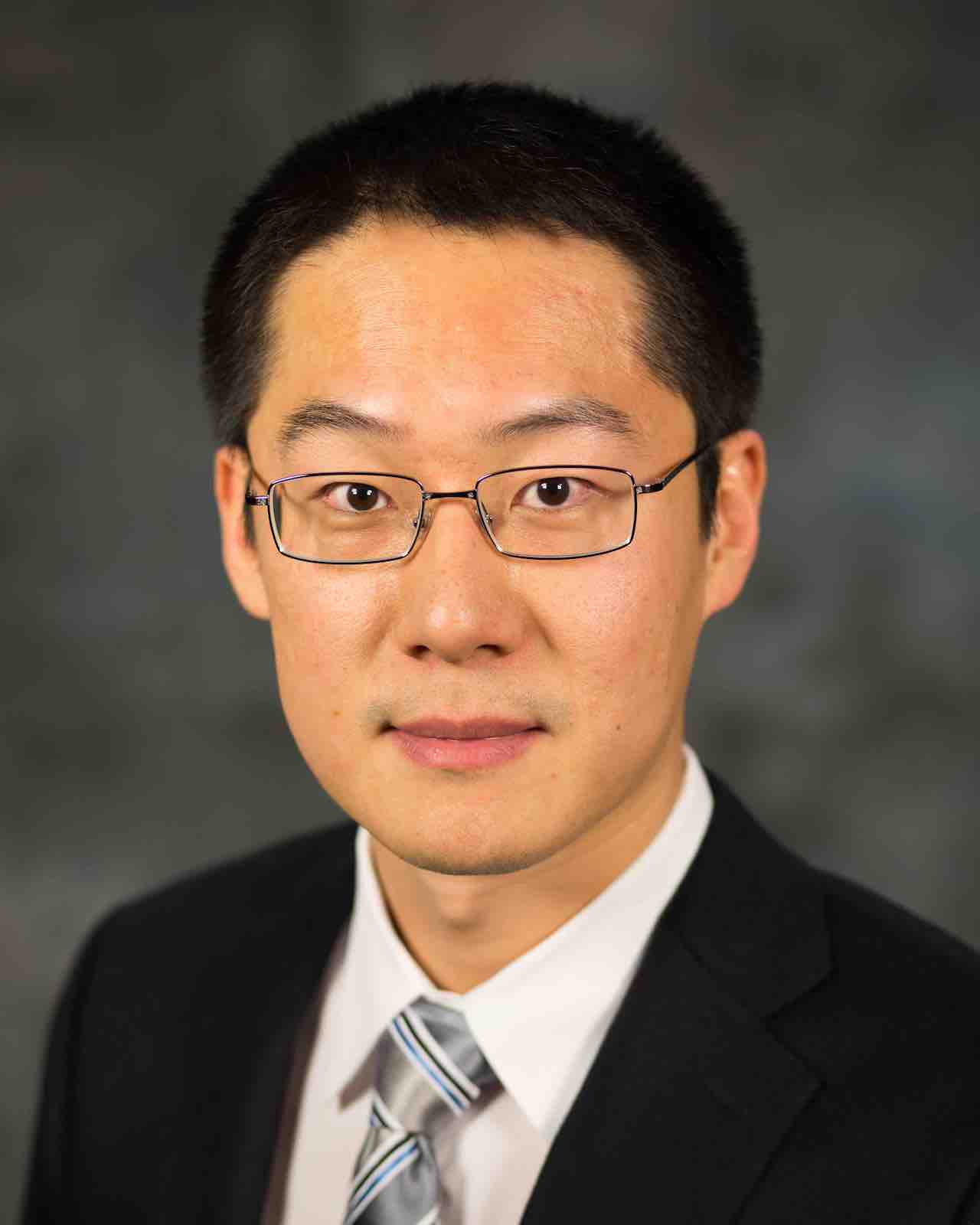}}]{Siyuan Liu} 
is Assistant Professor at Smeal College of Business, Pennsylvania State University. He received his first Ph.D. degree from Department of Computer Science and Engineering at Hong Kong University of Science and Technology, and the second Ph.D. degree from University of Chinese Academy of Sciences. 

His research interests include spatial and temporal data mining, social networks analytics, and data-driven behavior analytics.
\end{IEEEbiography}


% \newpage



%\vfill

% \enlargethispage{-1.67in}

\end{document}